\newcommand{\markas}[1]{\label{eq:#1}}
\newcommand{\helptext}[1]{}
\newcommand{\marginhelp}[1]{}

\newcommand{\gl}[1]{(\ref{eq:#1})}
\newcommand{\iek}{{\em i.e.}, }

\newcommand{\be}{\begin{equation}}
\newcommand{\ee}{\end{equation}}

\newcommand{\de}{\stackrel{\mbox{\tiny def}}{=}}
\newcommand{\refeq}[1]{\stackrel{(\ref{eq:#1})}{=}}

\newcommand{\Rthree}{{}^{(3)}R}
\newcommand{\gthree}{{}^{(3)}g}
\newcommand{\sign}{{\rm sgn}}

\newcommand{\biggraphics}[2]{\bf Figure omitted in the arXiv version.}
\newcommand{\rmd}{{\rm d}}
\newcommand{\flcommand}{}

\newcommand{\pacs}[1]{}
\newcommand{\submitto}[1]{}

\newcommand{\PR}{{\it Phys. Rev. }}
\newcommand{\PL}{{\it Phys. Lett. }}
\newcommand{\NC}{{\it Nuovo Cimento }}

\newcommand{\CQG}{{\it Class. Quantum Grav. }}

\documentclass[10pt,graphicx, a4paper]{article}
\usepackage{graphicx}
\usepackage{rotating}
\usepackage{subfigure}
\usepackage{amsfonts}

\begin{document}
\title{ \scshape A new approach to spherically symmetric junction surfaces
and the matching of FLRW regions}
\author{U Kirchner\\
\small Department of Mathematics and Applied Mathematics \\ \small
University of Cape Town\\ \small 7700 Rondebosch, South Africa \\
\small e-mail: uli@maths.uct.ac.za}
\date{\small 30. November 2003}
\maketitle


\begin{abstract}
We investigate timelike junctions (with surface layer)
between spherically symmetric solutions of the Einstein-field equation.
In contrast to previous investigations this is done in a coordinate
system in which the junction surface motion is absorbed in the
metric, while all coordinates are continuous at the junction surface.

The evolution equations for all relevant quantities are derived.
We discuss the no-surface layer case (boundary surface) and study the
behaviour for small surface energies. It is shown that one should
expect
cases in which the speed of light is reached within a finite proper time.

We carefully discuss necessary and sufficient conditions for a possible matching
of spherically symmetric sections.

For timelike junctions between  spherically symmetric space-time sections
we show explicitly that the time component of the Lanczos equation
always reduces to an identity (independently of the surface equation of state).

The results are applied to the matching of FLRW models. We discuss
`vacuum bubbles' and closed-open junctions in detail.
As illustrations several numerical
integration results are presented, some of them
indicate that the junction surface can reach the speed of light within a finite
time.

\end{abstract}

\pacs{04.20.Cv,98.80.Hw}
\submitto{\CQG}

\section{Introduction}

Recent measurements of the microwave background radiation support the idea that
our universe is highly isotropic and homogeneous \cite{spergel}. Cosmological
models with these properties are uniquely represented by the class of
Friedmann-Lema\^itre-Robertson-Walker (FLRW)
models. The observed flatness of the universe can then be explained by employing an
inflationary model \cite{guth}, which suggests an exponential expansion of the early universe
driven by a scalar field -- the inflaton field.

Nevertheless, it is often speculated that this might only be the local geometry,
while over a larger scale the universe is inhomogeneous and anisotropic, i.e.,
the matter content and geometry vary.
In particular, it appears as if the parameters necessary for life are highly
fine-tuned and in order to solve this ``fine-tuning problem'' it was suggested
that we live in one of many different FLRW regions -- most of them might be
unsuitable for life. The most prominent example is Linde's Chaotic Inflation Scenario \cite{linde1,linde2},
in which the different FLRW regions originate from
different almost homogeneous Planck-sized regions which experience a period of
exponential expansion.

When such models are discussed it is usually assumed that the transition region between
two almost FLRW regions is very small and can be approximated by a timelike junction hypersurface,
the so-called thin bubble wall. To find the motion of this hypersurface one has
to find the matching surface to the two solutions of the Einstein-field equation
representing the space-time on each side.

The matching conditions are of two different types. On the one hand there is the
purely geometric necessity that `things fit together' --- distances on the
junction surface
should have the same length when measured `inside' or `outside'. On a mathematical
level this reduces to a matching of the tangential metric components.

On the other hand there are the matching conditions which result from the assumed
validity of certain physical laws, in particular the energy-momentum conservation
across the junction surface and the validity of the Einstein-field equation on each side.
These conditions have been evaluated by C. Lanczos \cite{lanczos}, R. Dautcourt
\cite{dautcourt}, and in a ground-breaking work
by W. Israel \cite{israel}.

While these equations are in principle valid for any matching of two space-times satisfying
the Einstein-field equation, it is in practice impossible to handle their complexity
except for highly symmetric cases and for all practical applications
spherical symmetry is assumed. There is a vast amount
of literature, see e.g., \cite{israel,blau,berezin,lake}.

A good introduction to the standard method for dealing with spherically symmetric
junctions is given by K. Lake in \cite{lake}, where also
different matchings, including cosmological voids and vacuum bubbles, are studied
numerically. In \cite{berezin} V. A. Berezin, V. A. Kuzmin and I. I. Tkachev considered
the generic spherically symmetric case and expressed the relevant quantities
with invariants. Junctions arising from phase transitions in the early universe,
including junctions between sections of FLRW sections are discussed.

Our aim here is to present a new approach to junctions between spherically symmetric
space-times and to apply the formalism to junctions between FLRW models --- analytically
and numerically.

The approach will focus on the geometrical quantities describing the situation, i.e., the
distance of the junction surface from the centre of symmetry,
and not alone on the junction surface radius. In contrast to most other studies, we
do not evaluate the junction conditions (in particular the Lanczos equation) in
the original coordinate system or in Gaussian normal coordinates based on
the junction surface.
Instead we introduce new coordinates such that the junction surface is at
a fixed (new) ``radial``
coordinate and all coordinates are continuous at the junction surface. The motion of the
junction surface is now absorbed into the metric components.

In spherically symmetric cases the Lanczos equation has two non-trivial independent components
-- an angular  and a time component. While the first one leads to the well-known
evolution equations, there seems to be uncertainty about the interpretation
of the time component,
which is a second order (in time) differential equation for the junction surface motion.
It has been known that for certain particular cases this equation reduces to an
identity \cite{hajicek}.
Nevertheless, other authors\footnote{In particular this was suggested for the junction
between FLRW models.} suggested that this equation acts as a surface equation
of state \cite{berezin},
i.e., determines the surface pressure. Using the presented approach we will show that
the time component of the Lanczos equation is in fact an identity for all junctions
between spherically symmetric solutions of the Einstein-field equation.

It should be pointed out that there are special cases of junctions which could be
examined without employing junction conditions. If the $\gamma$-equation of state
and the cosmological constant have no discontinuity at the junction surface
then the spherically symmetric space-time can be described in terms of the
Lema\^itre-Tolman model. However, the really intersting question is how the
junction behaves if the inside and outside region have different dynamical
behaviour, i.e., different equation of state and cosmological constant. For these cases
one cannot avoid the use of junction conditions and all numerical examples given
in this paper will be of this kind.

Our approach differs in the use of a coordinate transformation
to minimize the number of discontinuous quantities and
to evaluate the standard junction conditions in a more convinient form
(leading to \gl{sl-matching-150} and \gl{sl-matching-200}). The
main results do agree with well known results in the literature ---
for example the evolution equation for the angular metric component
\gl{sl-matching-404} can be found in a similar (though
I believe less convinient) form in \cite{lake} and the results
for vacuum bubbles agree with findings in \cite{berezin} and \cite{sakai2}.

This paper will be structured as follows: In section \ref{sec-O3-matching} we re-examine
junction conditions for the matching of generic spherically symmetric sections. We
will re-derive the evolution and constraint equations using a new approach, based on a
coordinate transformation which makes {\em all} coordinates continuous at the
junction surface. We investigate the behaviour for small values of the surface-energy
density and discuss the special case of vanishing surface-energy density.
In section \ref{bw-sec-conditions}
we pay particular attention to constraints on the surface-energy density
and in section \ref{bw-sec-time-lanczos} we show that the time-component of the Lanczos equation is an identity.
This is followed by an application to the matching of FLRW sections
in section \ref{sec-FLRW-matching}
and three numerical examples in section \ref{bw-sec-numerics}.

\subsection{Notation}
We use standard notation and metric signature $(-+++)$.
Coordinates are labeled with greek indices, running from $0$ to $3$, where $0$ represents
the time coordinate. Latin indices label coordinates on the three-dimensional junction
hypersurface. As will be seen below, with our choice of coordinates the junction surface
is located at a fixed radial coordinate $R=1$, and hence latin indices take the values
$0,2$, and $3$ (or equivalently $t, \theta$, and $\phi$).

When considering quantities defined for certain hypersurfaces (like the junction surface)
it will be convenient to have a covariant derivative for this subspace. We will
use a vertical bar (like in $K_{a b |c}$) to refer to this covariant hypersurface derivative,
which is evaluated in the same way as above covariant derivative, but with all quantities
replaced by the corresponding hypersurface quantities. Any quantity referring to the
three-dimensional hypersurface of the junction surface will have a superscript `(3)', e.g.,
$\Rthree$ and $\gthree_{ab}$.

We will use a subscript $-$ or $+$ sign
to indicate whether quantities refer to  the in- or outside region, respectively.

\section{Matching of generic O(3)-symmetric sections}
\label{sec-O3-matching}
\marginhelp{sec-O3-matching}

\newcommand{\Norig}{{\mathfrak N}}
\newcommand{\torig}{\tau}

\subsection{The coordinate system}
\label{coord-sys}
\marginhelp{coord-sys}
Any spherically symmetric space-time (i.e., having O(3) symmetry) allows coordinates
such that the metric takes the form
\begin{equation}
\rmd s^2 = - \Norig^2(\torig,r)\rmd \torig^2 +l^2(\torig,r) \{\rmd r^2 + f^2(\torig,r) \rmd \Omega^2\},
\markas{m-coord-100}
\end{equation}
where $\Norig(\torig,r)$ is the so-called lapse function,
and $\rmd \Omega^2=\rmd \theta^2 + \sin^2(\theta)\rmd \phi^2$
the line-element on the two-dimensional unit-sphere.

We are interested in the matching of two spherically symmetric space-times, each
having a metric of the form \gl{m-coord-100}.
Generally, the coordinates will not match up at the junction surface and the
manifold is described by two different coordinate charts -- one for the inside
region (subscript $-$) and one for the outside region (subscript $+$).
 At any coordinate time (in the inside or outside region) the junction surface
itself is assumed to be a two-sphere which is described by its coordinate radius
in the inside and outside region, $\alpha_-(\torig_-)$ and $\alpha_+(\torig_+)$.

Since the inside and outside regions are originally described
by a metric in the form \gl{m-coord-100}
we will use the convention that a dot/prime refers to the proper time/radial derivative with respect
to the metric \gl{m-coord-100} at the junction surface, e.g.,
\[
\dot l_+ \de \frac{1}{\Norig_+} \frac{\partial}{\partial \torig} l_+(\torig,r) \Bigg|_{r=\alpha_+(\torig_+)}
\quad\mbox{ and } \quad
l_+' \de \frac{1}{l_+} \frac{\partial}{\partial r} l_+(\torig,r) \Bigg|_{r=\alpha_+(\torig_+)}.
\]

In order to describe the
motion of the junction surface one usually tries to find the evolution of the junction surface
radius in each coordinate system.
Here we want to suggest a different approach: we introduce a new coordinate system, such that
{\em all} the coordinates are continuous at the junction surface while only the transverse
metric components are discontinuous
at the junction surface.
The junction surface motion is now described by the evolution of the metric components.

This, however, should not be confused with Gaussian normal coordinates (which
are widely used in the discussion of junctions) as
our coordinate curves $x^\mu =$const. will be in general non-orhogonal at the
junction.

\paragraph{Constructing a continuous time coordinate}
Let us assume that the inside and outside spaces are given in terms of their metrics,
which take the form \gl{m-coord-100}. Generally the
time coordinates for the inside and outside region will not match up at the junction hypersurface.

Since the junction surface is timelike, each value of the time-coordiante (in
the inside or outside region) identifies a unique spherically symmetric
hypersurface of the junction. This establishes a
strictly monotonically increasing one-to-one relation between
the times on each side
$
\torig_+ = F(\torig_-).
$
Setting $\rmd \torig_- \de t$, $\rmd \tau_+ \de F'(t) \rmd t$
introduces the new continuous `global time coordinate' $t$.
This rescaling is independent of $r$ and leaves the form of the metric
\gl{m-coord-100} invariant. However,
if the original lapse function was constant then this re-scaling results in
a new time dependent lapse function which contains information about the junction surface
motion. For example, this will be the case for the matching of FLRW models.

If on the other hand the original lapse function $\Norig(\torig, r)$, where $\torig$ is the
original time coordinate, depends on the radial coordinate then
we can write the new lapse function (which makes the time coordinate continuous) as
\[
N(t,r)=F'(t) \Norig\left( \torig, r\right),
\]
where $\torig=\int^t_0 F'(t') \rmd t'$ and $t$  is the new time coordinate.
It is now $\eta(t)$ which contains information about the junction surface motion, while
$\Norig$ contains information about the background space. In particular, the
quantity
\[
\frac{N'(t,r)}{N(t,r)}=\frac{\Norig'(\torig,r)}{\Norig(\torig,r)}
\]
(taken as a function of proper time) does not depend on $F'$.
We will keep these issues in mind when we use the lapse function in the following calculations.

Finally it should be pointed out that there remains the `gauge freedom' to rescale
the new global time-coordinate $t$, which will be used later.

\begin{figure}[t]
\begin{center}
\includegraphics[width=5in]{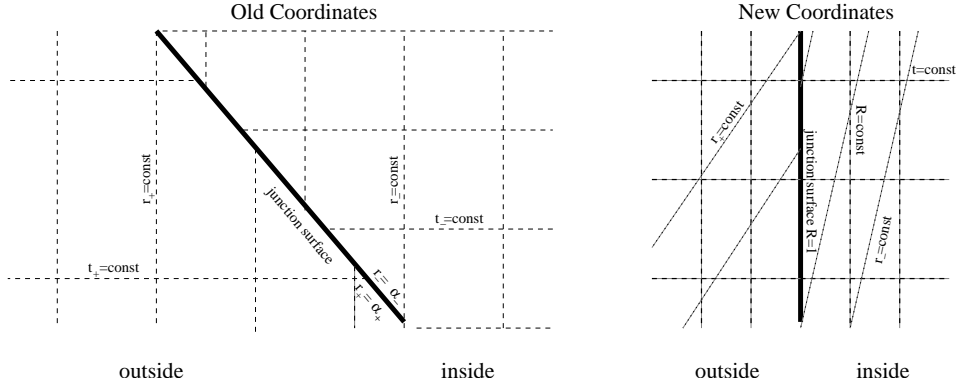}
\caption{
An illustration of the coordinate transformation. The original
radial coordinates $r_+$ and $r_-$ are rescaled such that
in the new coordinate system the junction surface is at a fixed `radial' coordinate
$R=1$. The time-coordinates are rescaled such that they match up at the
junction surface.
}\label{piii-fig5}
\end{center}
\end{figure}

\paragraph{Constructing a continuous radial coordinate}
Now we want to construct new radial coordinates such that the junction surface
is at a fixed radial coordinate $R=1$.
This can be achieved by setting $r_\pm=\alpha_\pm(t) R$, where the subscript $+$
refers to outside ($R>1$) and $-$ to the inside region, and $\alpha_\pm(t)$ is the coordinate
radius of the junction surface at time $t$.
The relation between the old and the new coordinates is illustrated in figure \ref{piii-fig5}.
With this new radial coordinate the metrics for the inside and outside region take the form
\be
\flcommand
\rmd s^2 = -N^2 \{1-R^2 \dot \alpha^2 l^2\} \rmd t^2
+ 2 \alpha N \dot \alpha l^2 R \rmd\!R \; \rmd t
+l^2\{\alpha^2 \rmd R^2 +f^2(\alpha R) \rmd \Omega^2\},
\markas{m-coord-200}
\ee
where a dot indicates the proper time derivative along paths
of constant $r,\theta,\phi$, i.e.,
$\dot \alpha \de \frac{1}{N}\frac{\partial}{\partial t}\alpha$.

\subsection{Geometric matching conditions}
At the matching surface $R=1$ the tangential metric components must be continuous. This gives us
the two matching conditions
\be
\left[
l f\right] =0.
\markas{O3-geom-cond-1}
\ee
and
\be
\left[
N^2(\dot \alpha^2 l^2 - 1)
\right]=0
,
\markas{O3-geom-cond-2}
\ee
where
\[
\flcommand
[g(R)]\de \lim_{R\rightarrow 1^+}(g(R))-\lim_{R\rightarrow 1^-}(g(R))
=\lim_{r_+\rightarrow \alpha_+(t)^+}(g_+(r_+))-\lim_{r_-\rightarrow \alpha_-(t)^-}(g_-(r_-))
.
\]
These relations identify two quantities which are continuous across the
junction surface and we define
\be
k(t,R) \de N \sqrt{1-\dot \alpha^2 l^2 R^2} \quad, \quad k\de k(t,1)=N\sqrt{1-\dot \alpha^2 l^2}\Big|_{R=1}
\markas{m-gma-200}
\ee
and
\be
w \de lf,
\markas{m-gma-300}
\ee
which are the tangential metric components. Note that $k(t,R)$ becomes complex for large
$\dot \alpha^2 l^2 R^2$. However, this is not of relevance for the problem at hand, since
we are only interested in the behaviour around the junction surface at $R=1$ where $k(t,R)$ is
real for timelike junction surfaces.

It follows from \gl{O3-geom-cond-2} that if the surface appears from one side as timelike,
it will so from the other side.
From now on let us assume that the junction surface is a timelike surface, \iek $\dot \alpha_\pm l_\pm <1$.

There is a remaining gauge freedom: we can rescale the global time coordinate, i.e.,
multiply the lapse functions with a time dependent factor.
One particularly useful choice is to rescale the time
such that the tangential metric component in the timelike direction parallel to the junction surface
becomes unity, i.e.,
\be
k = N \sqrt{1-\dot \alpha^2 l^2}\Big|_{R=1} =1.
\markas{m-gma-400}
\ee
To maintain generality we will not assume this choice until explicitly stated (in section
\ref{bw-subsec-surface-layer-matching}).

The two conditions \gl{O3-geom-cond-1} and \gl{O3-geom-cond-2}
are of pure geometric character -- they have to be
satisfied independently of the evolution equations at all times.
Taking the total derivative of \gl{m-gma-300} with respect to coordinate time we
obtain the corresponding restriction on the junction surface motion
\be
\frac{\rmd w}{\rmd t} = N\{ (lf)^\bullet + (lf)'\dot \alpha l \}
\qquad \left[ \frac{\rmd w}{\rmd t}\right]=0.
\markas{m-gma-600}
\ee
It should be noted that here $N$ is not independent, but depends via \gl{m-gma-400} on the
junction surface motion. Equations \gl{m-gma-400} and \gl{m-gma-600} generally have
two solutions for the junction surface
motion in terms of the surface radius evolution.

Let us note here that the metric of the timelike hypersurface representing the junction surface is given by
\[
\gthree_{\mu \nu} \rmd x^\mu \rmd x^\nu \de \rmd s^2_\Sigma = -k^2 \rmd t^2 + w^2 \rmd \Omega^2.
\]
The intrinsic geometry of the junction hypersurface is completely defined
by $k$ and $w$. Nevertheless, $w$ might not uniquely identify
the position of the junction surface.
Only if $l_+f_+$ {\em and} $l_-f_-$ are invertible functions of $\alpha_\pm$ then the position
of the junction surface is indirectly given by the value of $w$.

\subsection{Lanczos equation and Israel junction conditions}

Let us split the energy-momentum tensor in a regular and a $\delta$-function
part, so that
\be
T_{\mu \nu} = \delta(\eta) S_{\mu \nu} + \tilde T_{\mu \nu},
\markas{m-ic-100}
\ee
where $\tilde T_{\mu \nu}$ contains the regular part and $\eta$ is a function of the
coordinates which vanishes on the junction surface, is non-zero everywhere else,
and on the junction surface its gradient is a unit vector.
The tensor $S_{\mu \nu}$ is called the surface stress-energy (or energy-momentum) tensor. The
$\delta$-function restricts its influence to the junction surface and
we assume that it only depends on coordinates
on the junction surface, i.e., in our case this tensor does not depend on $R$.
The Lanczos equation \cite{lanczos} relates the
surface energy-momentum tensor $S_{\mu \nu}$ to the jump in the extrinsic curvature $K_{\mu \nu}$ of
the junction surface by
\be
\kappa S^{\mu \nu} = \gthree^{\mu \nu} [K] -[K^{\mu \nu}],
\markas{m-ic-200}
\ee 
or equivalently (after taking the trace and substituting back for $K$)
\be
\frac 1 \kappa [K^{\mu \nu}]=-S^{\mu \nu} +\frac 12 \gthree^{\mu \nu} S,
\markas{m-ic-300}
\ee
where $K \de K^\mu{}_\mu$, $S \de S^\mu{}_\mu$, and $\kappa \de 8\pi G$.
These two equations imply that the presence of a surface layer is equivalent
to a jump in the extrinsic curvature, \iek
$
\gamma^{\mu \nu} \de [K^{\mu \nu}] \ne 0.
$

To relate this conditions to the metric components on both
sides of the junction surface we have to find
expressions for the extrinsic curvature. We start with the normal to the junction surface,
which is given by
\be
n_\mu = \delta_\mu^R \frac{l\alpha }{\sqrt{1-R^2\dot \alpha^2 l^2}} \Bigg|_{R=1},
\markas{m-ic-500}
\ee
and the unique timelike unit-vector tangential to the junction surface
and orthogonal to the spherical symmetric
subspace, which is given by
\be
u^\mu=
\frac{1}{k} \delta^\mu_t.
\markas{m-ic-600}
\ee
The extrinsic curavure $K_{\mu \nu} = n_{(\lambda;\kappa)} h^\lambda{}_\mu h^\kappa{}_\nu$,
where $h^{\mu \nu} = g^{\mu \nu} - n^\mu n^\nu$ is the projection tensor onto
the junction surface, has the two independent components
\be
K_{\mu \nu} u^\mu u^\nu = \frac{1}{k^3 N l \alpha} \Gamma_{Rtt}
=-\frac{\dot N + 2N'l\dot \alpha + \dot \alpha^2 l^2 N (\dot l/l)}{k^3 \dot \alpha l}.
\markas{m-ic-1300}
\ee
\be
K^{\theta}{}_{\theta} = \frac N{k w} (\dot \alpha l (lf)^\bullet + (lf)'),
\markas{m-ic-1400}
\ee
where we used
the coordinate time derivative
of \gl{m-gma-400} to simplify the first expression.

This form of the extrinsic curvature implies that the surface
energy-momentum tensor $S^{\mu \nu}$ is
diagonal and of perfect-fluid form (in the three-dimensional hypersurface space).
We introduce the surface energy density $\rho_s$ and pressure $p_s$ such that
\be
S^{\mu \nu} = (\rho_s + p_s) u^{\mu} u^{\nu} + p_s \gthree^{\mu \nu}.
\markas{m-ic-1500}
\ee
The Lanczos equations \gl{m-ic-300} are now given by the two equations
\begin{eqnarray}
\left[
N\{
\dot \alpha l (fl)^\bullet +(fl)'
\}
\right]
&=& -\frac \kappa 2 \rho_s w k \markas{m-ic-1900}\\
\left[
\dot N + 2N'l\dot \alpha + \dot \alpha^2 l^2 N (\dot l/l)
\right]
&=& \kappa \left( \frac 12 \rho_s +p_s\right) k^3.
\markas{m-ic-2000}
\end{eqnarray}

The Gauss-Codazzi equations relate the curvature at one point to the
extrinsic and intrinsic curvature of a hypersurface which passes through this
point
by
\[
R_{\mu \nu \lambda \kappa} e_a{}^\mu e_b{}^\nu e_c{}^\lambda e_d{}^\kappa
= {}^{(3)}R_{abcd} + \epsilon(K_{ad}K_{bc}-K_{ac}K_{bd})
\]
and
$ R_{\mu \nu \lambda \kappa} n^\mu e_b{}^\nu e_c{}^\lambda e_d{}^\kappa
= K_{bc|d} - K_{bd|c}$,
where $e_a{}^\mu$ is a coordinate basis of the junction surface,
a vertical bar denotes the covariant derivative with respect to
the induced hypersurface metric $h_{\mu \nu} = g_{\mu \nu} - n_\mu n_\nu$,
and $\epsilon \de n^\mu n_\mu$ equals $+1$ for timelike hypersurfaces, and
$-1$ for spacelike hypersurfaces.
For a timelike hypersurface these equations lead to \cite{misner-thorne-wheeler}
(after substituting \gl{m-ic-300})
\be
[G_{\mu \nu} e_a{}^\mu n^\nu] = -\kappa S_a{}^b{}_{|b}
\markas{m-gc-600}
\ee
\be
[G_{\mu \nu} n^\mu n^\nu] = \frac 12 [K^2-K_{\mu \nu}K^{\mu \nu}]
= \kappa S^{\mu \nu} \bar K_{\mu \nu},
\markas{m-gc-700}
\ee
where $\bar K_{\mu \nu} \de \frac 12 (\lim_{R \rightarrow 1^+}K_{\mu \nu}
+ \lim_{R \rightarrow 1^-}K_{\mu \nu})$ and
$G_{\mu \nu}\de R_{\mu \nu}-\frac 12 g_{\mu \nu} R$ is the Einstein tensor.

The second of these equations represents
nothing else than the definition of the surface stress-energy tensor $S^{\mu \nu}$
(which was substituted) and hence is redundant.
Using \gl{m-ic-1500} to evaluate \gl{m-gc-600} we find that the only non-vanishing
component is the time component, which is given by (using $e_0{}^\mu=u^\mu$)
\be
k u^\mu {\rho_s}_{;\mu} + k(\rho_s + p_s) u^a{}_{|a} = \kappa^{-1} [G_{\mu \nu} u^\mu n^\nu]
=[T_{\mu \nu} u^\mu n^\nu],
\markas{m-gc-800}
\ee
where the last equality follows from the Einstein equation and $u^a{}_{|a}=u^\mu (2 \ln (lf))_{;\mu}$.
Together with an equation of state for the surface energy and pressure densities this
equation describes the evolution of the `matter' on the surface.

\subsection{Matching with surface-layer}
\label{bw-subsec-surface-layer-matching}
\marginhelp{bw-subsec-surface-layer-matching}

The second geometric matching condition shows that $k\de N \sqrt{1-\dot \alpha^2 l^2}$
is continuous across the junction surface. Hence we can express the junction surface motion
in terms of the lapse function by
\be
\dot \alpha l = \pm \sqrt{1- \left( \frac kN \right)^2}.
\markas{sl-matching-100}
\ee
For convenience we will choose now $k=1$, i.e., the coordinate time
corresponds to proper time along the curves $R=1,\theta,\phi$ constant on the junction surface.
Setting $u\de N/k$ and $j_\pm \de \sign(\dot \alpha_\pm)$
the angular component of the extrinsic curvature \gl{m-ic-1400}
on both sides of the junction surface and
equation \gl{m-gma-600},
the derivative of the junction surface radius with respect to coordinate time,
are given by
\begin{eqnarray}
K^\theta{}_\theta&=& (uw' + j \dot w\sqrt{u^2-1})/w
\markas{sl-matching-150}\\
L &\de& \frac{\rmd w}{\rmd t} = u\dot w + j w' \sqrt{u^2-1}
\markas{sl-matching-200}
\end{eqnarray}
A valid matching between two spherically symmetric sections, each satisfying
the Einstein-field equations, must satisfy the two geometric matching conditions
(\gl{O3-geom-cond-1} and \gl{O3-geom-cond-2}) and the two independent components
of the Lanczos equation (\gl{m-ic-1900} and \gl{m-ic-2000}). We first note, that
if the first geometric
matching condition (matching of the surface radius) \gl{O3-geom-cond-1} is satisfied
initially, then it is sufficient to demand that its coordinate time derivative
\gl{m-gma-600} is satisfied at all times.
Secondly, with our choice of variables the second geometric matching condition
\gl{O3-geom-cond-2} is nothing more than an identity --- with \gl{sl-matching-100}
it has already been used to eliminate one variable.
Thirdly, as will be shown in section \ref{bw-sec-time-lanczos},
equation \gl{m-ic-2000}, the time-component of the Lanczos
equation, is in fact identically satisfied if all the other matching conditions
are satisfied, the Einstein-field equations are valid on each side,
and the surface-matter evolution is given by \gl{m-gc-800}.

We conclude that the matching conditions are completely represented by \gl{m-gma-600}, 
the coordinate time derivative of the first geometric matching condition,
and the angular component of the Lanczos equation \gl{m-ic-1900} together with
an initial matching of the proper surface radius $w=lf$. With
our choice of variables these equations take the (surprisingly symmetric) form
\begin{eqnarray}
[L] &=& [u\dot w + j w' \sqrt{u^2-1}]=0 \\
\
[w K^\theta{}_\theta ] &=& [uw' + j \dot w\sqrt{u^2-1}]=-E,
\markas{sl-matching-300}
\end{eqnarray}
where $w\de lf$ is the proper surface radius of the spherical junction surface on
each side and
\be
E \de \frac{\kappa \rho_s w}{2}
\markas{sl-matching-400}
\ee
quantifies the energy-content of the layer.
To find a relation between $K^\theta{}_\theta$ and $L$ we square \gl{sl-matching-150}
and substitute $L^2$ from the square of \gl{sl-matching-200} and obtain
\be
(w K^\theta{}_\theta)^2=L^2 + a,
\markas{sl-matching-402}
\ee
where $a\de w'^2-\dot w^2$.
Versions of this equation have been given in \cite{berezin} and \cite{sakai2}.
For $E\ne 0$ we can express $K^\theta{}_\theta$ in terms of $L$ by using an algebraic identity as
\be
wK^\theta{}_\theta = \frac{[(wK^\theta{}_\theta)^2]\pm [wK^\theta{}_\theta]^2}{2 [wK^\theta{}_\theta]}
=\frac{b \pm E^2}{-2E},
\markas{sl-matching-403}
\ee
where $b\de a_+-a_-$. The explicit expression for $L$ in terms of $E$
takes the form
\be
\flcommand
L^2 = \left( \frac{b -E^2}{2E} \right)^2 - a_-
= \left( \frac{b +E^2}{2E} \right)^2 - a_+
=(E^4-2E^2(a_+ + a_-)+b^2)/4E^2
,
\markas{sl-matching-404}
\ee

Note that we find by differentiating \gl{sl-matching-150} and \gl{sl-matching-200}
with respect to $u$ (taking $w, \dot w,$ and $w'$ to be independent of $u$) the helpful relations
\be
j \sqrt{u^2-1} \frac{\partial L}{\partial u} = w K^\theta{}_\theta
\quad \mbox{and} \quad
j \sqrt{u^2-1} \frac{\partial wK^\theta{}_\theta}{\partial u}=L,
\markas{sl-matching-405}
\ee
which are valid on each side of the junction surface.

The geometric matching condition \gl{sl-matching-200} can be solved for
$u_\pm$ and $j_\pm$ in terms of the time derivative of the surface radius $L$
and the extrinsic curvature component $K^\theta{}_\theta$
\be
u =
\frac{\dot w L - w'wK^\theta{}_\theta}{-a}.
\markas{sl-matching-410}
\ee
We note that differentiating \gl{sl-matching-410} with respect to $u$ and using
\gl{sl-matching-405} yields
\be
j \sqrt{u^2-1}=\frac{w'L-\dot w wK^\theta{}_\theta}{a},
\markas{sl-matching-430}
\ee
what also determines the sign of $j$ and hence
the radial direction of motion of the junction surface for each side.

\subsection{The `no surface-layer' case}
\label{matching-bs} \marginhelp{matching-bs}

If all tangential components of the extrinsic curvature are continuous
at the junction surface then it follows from \gl{sl-matching-300}
that the surface-energy density $\rho_s$ vanishes. In this case
the junction hypersurface is called a boundary-surface.
It is an immediate consequence of \gl{sl-matching-402} that
\[
E=\frac{\kappa \rho_s w}{2}=0
\Rightarrow
b = [a] =[{w'}^2-{\dot w}^2]= 0,
\]
and $b=0 \Rightarrow E=0 $ or $E=wK^\theta_{- \theta}/2=-wK^\theta_{+ \theta}/2$.

Feasible solutions need to satisfy the two geometric matching
conditions \gl{O3-geom-cond-1} and \gl{O3-geom-cond-2}, the derivative
of the first matching condition \gl{m-gma-600} and the matching of
the extrinsic curvature \gl{m-ic-1900}. Recognizing the similar
structure of \gl{m-gma-600} and \gl{m-ic-1900} we
form two new equivalent equations by adding and subtracting
the two equations. The result reads
\[
[N(1-\dot \alpha l)(w'-\dot w)]=0
\qquad
[N(1+\dot \alpha l)(w'+\dot w)]=0.
\]
Using the factorized form of the second geometric matching condition \gl{O3-geom-cond-2}
and defining $q=N(1-\dot \alpha l)$ this becomes
\be
[q(w'-\dot w)]=0 \qquad [\frac 1 q (w'+\dot w)]=0.
\markas{nsl-0030}
\ee
If both $w'-\dot w$ and $w'+\dot w$ vanish separately on both sides, then the system
becomes an identity and the junction surface motion remains undefined. Let us
assume now that this is not the case.

We note that for $[w']=[\dot w]=0$ the system is solved for
any $q_+=q_-$. Hence, if the angular component of the metric and its first order
proper time and radial derivatives are continuous, then the junction surface motion
does not follow from the matching conditions. In particular, this is the case
for the trivial matching of two identical space-times, were we have an
`imaginary junction surface', which could be placed anywhere.

Let us from now on assume that at least one of the proper derivatives of $w$
is not continuous at the junction surface.

If $w' \pm \dot w$ is zero on one side, it has to be zero on
the other side too (otherwise no matching is possible) and
one of the equations \gl{nsl-0030} is identically satisfied.

However, if all $w'_\pm \pm \dot w_\pm$ are non-zero then the condition $b=0$
also guarantees that the two linear equations \gl{nsl-0030} are
linearly dependent. The motion of the boundary surface is then described
by the four equations (if $w' \pm \dot w$ is zero then one
of the options in the first equation is undefined, but the other option
is then still valid)
\be
\flcommand
\frac{N_+(1-\dot \alpha_+^2 l_+^2)}{N_-(1-\dot \alpha_-^2 l_-^2)}
= \frac{w'_- - \dot w_-}{w'_+ - \dot w_+}
= \frac{w'_+ + \dot w_+}{w'_- + \dot w_-}
\qquad
N^2_\pm (1- \dot \alpha_\pm^2 l_\pm^2) = 1
\qquad
\frac{\rmd b}{\rmd t} =0,
\ee
where
\be
\frac{\rmd b}{\rmd t} =2 [N w' \{ (w')^\bullet + w'' \dot \alpha l \}
-\dot w N\{\ddot w + (\dot w)' \dot \alpha l\}].
\ee

\subsection{Expansion for small surface-energy densities}
\label{bw-subsec-expansion}
\marginhelp{bw-subsec-expansion}

As will be seen in the numerical examples given later, in many cases the
surface-energy density (and hence $E$) approaches zero at some finite coordinate time. In the case of $b\ne 0$
the dynamic quantities can be approximated by a series expansion in terms of $E$. We start
by re-writing the exact expression for the extrinsic curvature \gl{sl-matching-403} as
\[
wK^\theta_{\pm \theta} = -\frac{b}{2E} \mp \frac E2.
\]
It follows then from \gl{sl-matching-404} that
\[
\pm L=\frac{|b|}{2E} - \frac{a_++a_-}{2|b|}E + O(E^3),
\]
where $O(E^3)$ represents terms of the order $E^3$ or smaller. Furthermore, from \gl{sl-matching-410} we
find for $u=N/k$ the expansion
\[
\flcommand
u_\pm = -\frac{w'_\pm b + \sign(L) \dot w_\pm |b|}{2a_\pm} \frac{1}{E}
+\frac{\sign(L) \dot w_\pm(a_++a_-)/|b| \mp w'_\pm}{2a_\pm} E +O(E^3).
\]
As the surface-energy density approaches zero the lapse functions (given by $u_+$ and $u_-$)
diverge and the proper speed of the junction surface approaches the speed of light
quadratically since
\[
|\dot \alpha_\pm l_\pm|=1-\frac{2 \alpha_\pm l_\pm^2}{w_\pm'b+\sign(L) \dot w_\pm |b|}E^2 +O(E^4).
\]
To examine if and how $E$ approaches zero we finally expand the evolution equation \gl{m-gc-800}
\[
\frac{\rmd E}{\rmd t} = \frac{\kappa}{2} \left(
w [T_{\mu \nu} u^\mu n^\nu] - \frac{|b|}{w}\left(\gamma_s-\frac 12\right)
\right) +O(E^2).
\]
Generally the first term diverges as we approach the speed of light --- for example in the case
of a perfect fluid one finds 
\[
|T_{\mu \nu} u^\mu n^\nu| = \left( \frac{w'b+\sign(L) \dot w |b|}{2a}\right)^2 (\rho + p)\frac{1}{E^2}
+O(E^0).
\]
We conclude that if the energy-momentum contribution $[T_{\mu \nu} u^\mu n^\nu]$ has a sign opposite
to $E$, then  $E$ accelerates towards zero. In many cases $E$ will reach zero at some
finite coordinate time $t_0$. Close to this point and assuming that the time dependence of
all other terms is negligible we have $E \propto (t_0-t)^{1/3}$. This implies that
the lapse functions are integrable and the junction surface reaches the speed of light
(on each side) within a finite proper time. Here our formalism breaks down and one would
need a separate treatment of these singular cases. We want to speculate here that
at these points the junction surface turns spacelike. 

On the other hand, if the sign of $[T_{\mu \nu} u^\mu n^\nu]$ is the same as the sign of $E$
then $E$ cannot get arbitrarily close to zero. In some cases (see figure \ref{bw-numeric-graph13}) $E$ will
oscillate around some value (which is itself time dependent). Even in these cases we can
encounter divergencies resulting from diverging $a_\pm$ and $b$. This can lead to
non-integrable lapse functions - from each side the junction seems to exist forever,
but an observer who moves along the junction encounters a singular point after a finite time.
At this point the surface energy density is zero and again the formalism breaks down.

It should be noted that in the case of a perfect fluid on both sides of the junction
the sign of the stress-energy contribution $[T_{\mu \nu} u^\mu n^\nu]$ depends on the
energy density $\rho$ and pressure $p$
on both sides, i.e., on the equations of state. Hence whether a particular junction reaches the
speed of light within a finite time or not might depend on the equation of state on each side.
In section \ref{bw-sec-numerics} we give a numerical example for such a case.

Because points on the junction surface are not causally connected
a spacelike junction surface has a very different physical interpretation . In such cases
the junction surface cannot be treated as an `evolving system' on its own, but
rather as some kind of (spacelike) transition surface which is generated by the physics underlying
the cosmological model.

Usually a timelike junction surface is used to model the time evolution of a spatially {\em localized}
inhomogeneity.
If a junction surface turns spacelike a breakdown in the thin wall approximation
must have occurred.

\section{Necessary and sufficient conditions for a possible matching}
\label{bw-sec-conditions}
\marginhelp{bw-sec-conditions}

\subsection{Demanding real solutions for $L$}
\begin{figure}[ht]
\begin{center}
\includegraphics[width=3.5in]{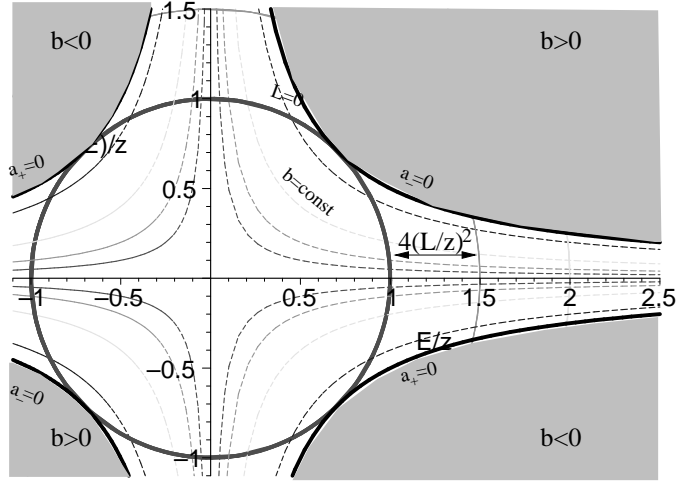}
\caption{
\helptext{bw-fig4new}
For the case $a_+,a_- \ge 0$ this figure shows the
region in the $\frac E z - \frac b{Ez}$-plane with $z \de \sqrt{2(a_+ + a_-)}$
where there is no solution for the junction surface motion. The dashed lines show
the curves of constant $b$, the circles are the lines of constant $L/z$, and the shaded
region is not feasible since by construction $2|b| \le z^2$.}
\label{bw-fig4new}
\end{center}
\end{figure}

From \gl{sl-matching-404} we find with $L^2\ge 0$ a necessary condition for the
existence of solutions which restricts the allowed values for $E$, such that
\be
E^4-2E^2(a_++a_-)+b^2 \ge 0.
\markas{sl-matching-480}
\ee
The roots of the quadratic polynomial (in $E^2$) on the left-hand side are given by
$a_++a_-\pm2\sqrt{a_+a_-}$. For $a_+ a_- <0$
or $a_+<0, a_-<0$
all values of $E$ are feasible.
Hence it is only for $a_+ \ge 0,a_-\ge 0$ that restrictions on $E$ arise, which take
the form
\be
0\le |E| \le |\sqrt{a_+}-\sqrt{a_-}| \quad \mbox{or} \quad \sqrt{a_+}+\sqrt a_- \le |E|.
\markas{sl-matching-490}
\ee
The shape of the forbidden region takes a particular simple form in
the $E-\frac b E$ plane, which is illustrated
in figure \ref{bw-fig4new}.

The two disjoint regions allowed for $|E|$ given by \gl{sl-matching-490}
are easily distinguished by $E^2 \le |b|$ and $E^2 \ge |b|$.
Furthermore, from \gl{sl-matching-403} we find
$\sign(K^\theta_{- \theta})=\sign(K^\theta_{+ \theta})=-\sign(b/E)$
for $E^2< |b|$ and $\sign(K^\theta_{-\theta})=\sign(E),\sign(K^\theta_{+\theta}) =-\sign(E)$ for $E^2> |b|$.

\subsection{Proper time relations}
By setting $k=1$ it follows from  \gl{m-gma-200} the condition
\be
N=u \ge +1,
\markas{sl-matching-495-100}
\ee
i.e., on each side of the junction surface proper time must proceed faster
(with respect to the time coordinate) than on the junction surface.
To investigate the resulting constraints on the surface energy density
we start by noting that
\be
|w'(wK^\theta{}_\theta)-a|-|\dot w L|
\left\{ \matrix{\ge 0 \mbox{ for } a\ge0\cr \le 0 \mbox{ for } a \le 0} \right.,
\markas{sl-matching-495-200}
\ee
which can be easily verified by squaring and substituting from \gl{sl-matching-402}.
Substituting \gl{sl-matching-410} the inequality \gl{sl-matching-495-100} takes
the form
\[
\frac{(w'(wK^\theta{}_\theta)-a)-\dot w L}{a} \ge 0.
\]
It follows from \gl{sl-matching-495-200} that for $a > 0$ we need
$
w'(wK^\theta{}_\theta)-a \ge 0,
$
while for $a<0$
\be
\dot w L \ge 0.
\markas{sl-matching-495-500}
\ee

\begin{figure}
{
\centering
\subfigure[Allowed ranges for $E$ if $\epsilon>0$.]{\label{bw-fig-E-ranges_a2}
\includegraphics[width=2in]{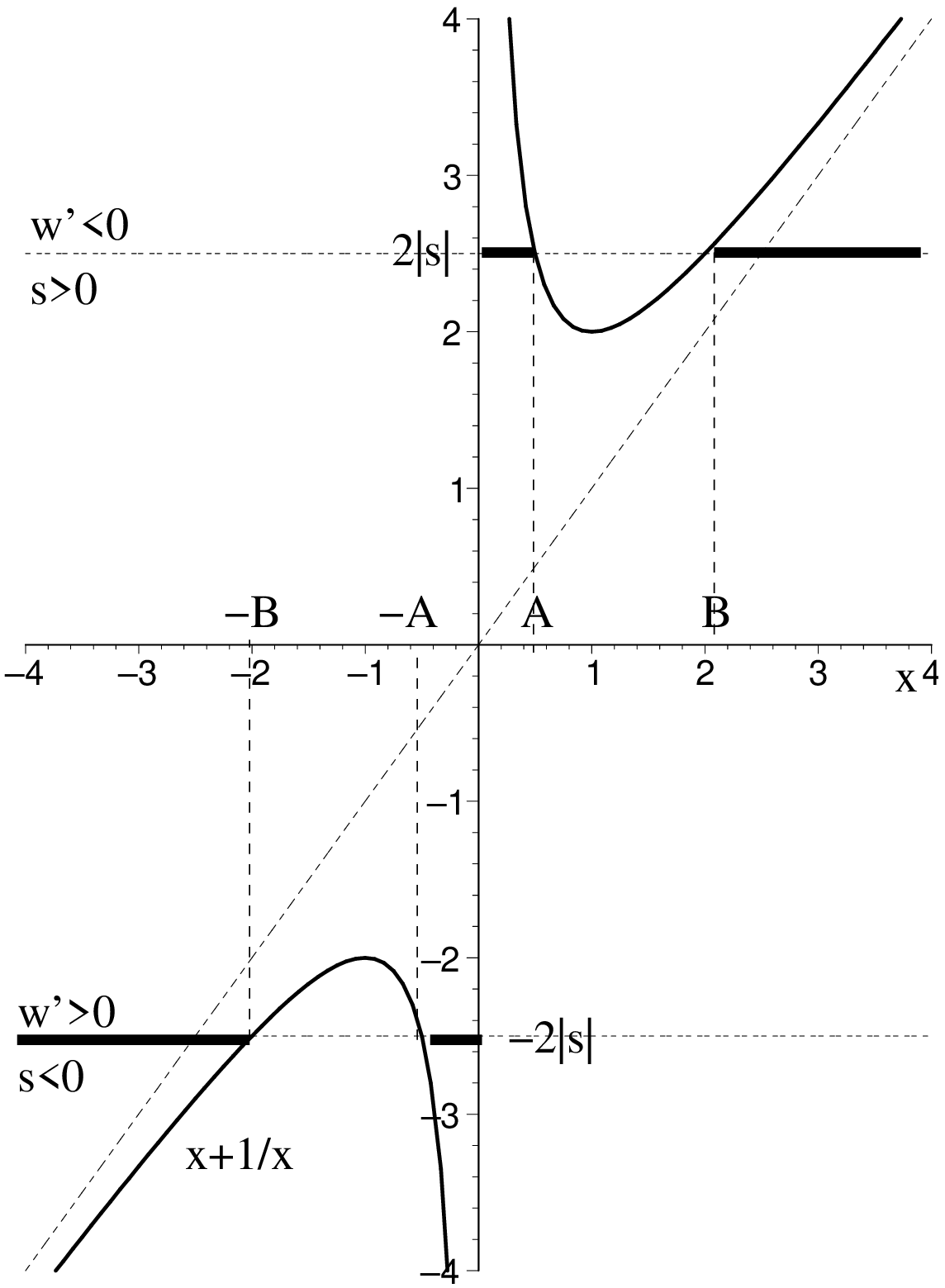}}
\hspace{.1cm}
\subfigure[Allowed ranges for $E$ if $\epsilon<0$.]{\label{bw-fig-E-ranges_a1}
\includegraphics[width=2in]{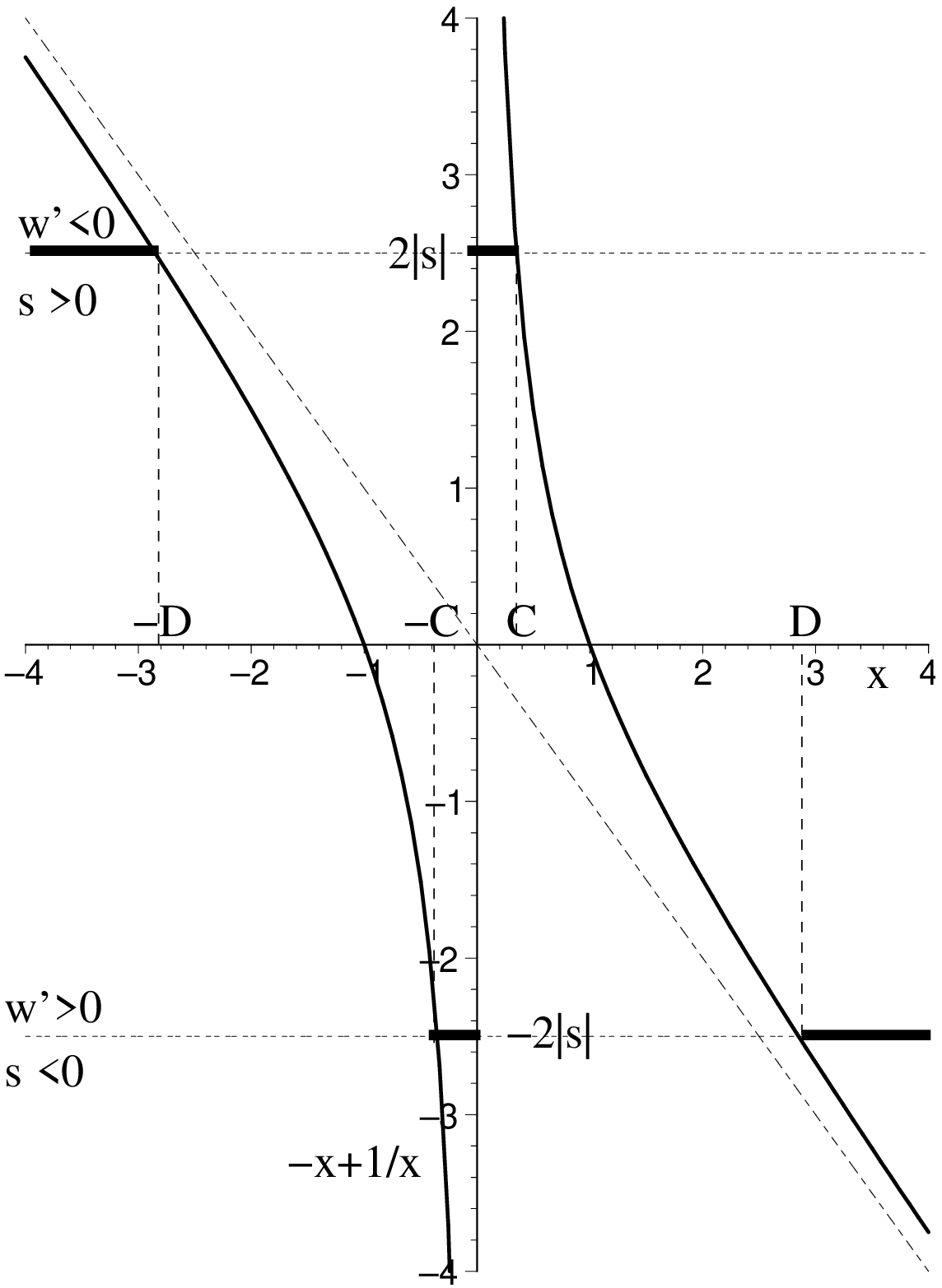}}
\caption{Visualization of the inequality \gl{sl-matching-495-700} for $\epsilon>0$
(figure \ref{bw-fig-E-ranges_a2}) and $\epsilon<0$
(figure \ref{bw-fig-E-ranges_a1}). The lower and upper horizontal line correspond to
$w'_\pm >0$ and $w'_\pm<0$, respectively. Here $s_\pm \de -a_\pm/(w'_\pm\sqrt{|b|})$ are the
terms on the right-hand side of \gl{sl-matching-495-700}. The allowed
ranges are indicated by a bold line. The end-points are given by ($A$ and $B$ are only
defined for $|s|>1$)
$
A = |s|-\sqrt{s^2-1},
B = |s|+\sqrt{s^2-1},
C = -|s|+\sqrt{s^2+1},
D = |s|+\sqrt{s^2+1}$.
}
\label{bw-fig-E-ranges}
}
\end{figure}

Let us first consider the case $a>0$. Using \gl{sl-matching-403} the condition becomes
$
w'(b + \sigma E^2)/(-2E) -a \ge 0,
$
where $\sigma=\pm 1$ corresponding to the outside ($+$) and inside ($-$) case.
In the following let us use the convention that if $a,w$ refer to the quantities
on one side, then $a_*,w_*$ refer to the quantities on the other side of the
junction.
By setting $x \de \sign(b)E/\sqrt{|b|}$, $\epsilon_\pm \de \pm \sign(b)=\sign(a-a_*)$,
and
$
s_\pm \de -\frac{a_\pm}{w'_\pm \sqrt{|b|}}
$
we bring the inequality in the form
\be
\frac 1 x + \epsilon_\pm x \left\{ \matrix{\le 2 s_\pm \mbox{ for } w'_\pm>0\cr
\ge 2 s_\pm \mbox{ for } w'_\pm<0}
\right.
.
\markas{sl-matching-495-700}
\ee
The allowed ranges for $x$ are illustrated in figure \ref{bw-fig-E-ranges}.

\paragraph{The case of $\epsilon>0$}
Let us note that if $\epsilon_+>0$ then $\epsilon_-<0$ and vice versa ($\epsilon_*<0$).
The sign of the surface energy density is now
determined by
\be
\sign(\rho_s) = - \sign(b) \sign(w').
\markas{bw-eq-constraints-600}
\ee
Furthermore, if $|s|=|a/(w' \sqrt{|b|})|\le 1$ then no restrictions are placed
on $|x|$ (but further restrictions could come from $a_*>0$).
For $|s|>1$ the allowed range can be found by setting $|x|=e^z$ and hence
$|1/x+x|=2 \cosh(z)$. Using $\cosh^{-1}|s|=\ln(|s|+\sqrt{s^2-1})$ one
obtains the ranges
\[
0 < |x| \le
|s|-\sqrt{s^2-1}
\quad \mbox{ or } \quad
 |s| + \sqrt{s^2-1} \le |x|.
\]

\paragraph{The case of $\epsilon<0$} Similarly to the last case, if $\epsilon<0$ then
$\epsilon_*>0$. Hence each case will occur once at the junction.
A similar procedure as above yields the restrictions
\[
\flcommand
|s|-\sqrt{s^2+1} \le \sign(w') x <0
\quad \mbox{ or } \quad
|s|+\sqrt{s^2+1} \le \sign(w')x.
\]

\paragraph{The case of $a>0$ on both sides}
If $a$ is positive on both sides then $x$ can only take values which lay in the
intersection of the allowed ranges on each side. The allowed intervals
differ, depending on the signs of $w'_+$ and $w'_-$. All possible
cases are shown in table \ref{bw-table-E-ranges}.
We note that in every case the allowed values for $E$ have the same
sign.

As a particularly important case (for the matching of FLRW models) and an illustrative
example we evaluate the restrictions on the surface energy density for $a_+,a_->0$ and
$w'_+,w'_->0$. In this case $x$ has to be negative and hence
$
\sign(\rho_s)=\sign(E)=-\sign(b).
$
If one assumes on physical grounds that $\rho_s$ should be positive then no
matching will be possible if $b$ is positive.

\begin{table}
\footnotesize
\begin{center}
\begin{tabular}{|l|c|c|}
\hline
&$\matrix{b>0\cr\Rightarrow \epsilon_+=+1,\epsilon_-=-1}$
&$\matrix{b<0\cr\Rightarrow \epsilon_+=-1,\epsilon_-=+1}$\\
\hline
$\matrix{w'_+>0\cr w'_->0}$
&$\matrix{\sign(\rho_s)=-1\cr\sqrt{|b|} \max(-A_+,-C_-)\le E < 0
\mbox{ for } |s_+|>1\cr
-\sqrt{|b|} C_-\le E < 0
\mbox{ for } |s_+|\le 1
}$
&$\matrix{\sign(\rho_s)=+1\cr0<E\le \sqrt{|b|} \min(A_-,C_+)\mbox{ for }|s_-|>1\cr
0<E\le \sqrt{|b|} C_+\mbox{ for }|s_-|\le 1}
$\\
\hline
$\matrix{w'_+>0\cr w'_-<0}$
&$\matrix{\sign(\rho_s)=-1\cr E \le \sqrt{|b|} \min(-B_+,-D_-)
\mbox{ for } |s_+|>1\cr
E \le -\sqrt{|b|} D_- \mbox{ for } |s_+|\le 1
}$
&$\matrix{\sign(\rho_s)=-1\cr E \le \sqrt{|b|} \min(-B_-,-D_+) \mbox{ for } |s_-|>1\cr
E \le -\sqrt{|b|} D_+ \mbox{ for } |s_-|\le 1
}$\\
\hline
$\matrix{w'_+<0\cr w'_->0}$
&$\matrix{\sign(\rho_s)=+1\cr \sqrt{|b|} \max(B_+,D_-)\le E \mbox{ for } |s_+|>1\cr
\sqrt{|b|} D_- \le E \mbox{ for } |s_+|\le 1
}$
&$\matrix{\sign(\rho_s)=+1\cr \sqrt{|b|} \max(B_-,D_+)\le E \mbox{ for } |s_-|>1\cr
\sqrt{|b|} D_+\le E \mbox{ for } |s_-|\le 1
}$\\
\hline
$\matrix{w'_+<0\cr w'_-<0}$
&$\matrix{\sign(\rho_s)=+1\cr 0<E \le \sqrt{|b|} \min(A_+,C_-) \mbox{ for }|s_+|>1\cr
0<E \le \sqrt{|b|} C_- \mbox{ for }|s_+|\le 1
}$
&$\matrix{\sign(\rho_s)=-1\cr \sqrt{|b|} \max(-A_-,-C_+)\le E <0 \mbox{ for }|s_-|>1\cr
-\sqrt{|b|} C_+\le E <0 \mbox{ for }|s_-|\le 1
}$\\
\hline
\end{tabular}
\caption{\helptext{bw-table-E-ranges}For $a_+>0$ and $a_->0$ this table shows
the allowed region for $E$ for all possible combinations of $\sign(w'_+)$ and
$\sign(w'_-)$. Here
$
A_\pm = |s_\pm|-\sqrt{s_\pm^2-1}\le 1; B_\pm = |s_\pm|+\sqrt{s_\pm^2-1}\ge 1;
C_\pm = -|s_\pm|+\sqrt{s_\pm^2+1}\le 1; D_\pm = |s_\pm|+\sqrt{s_\pm^2+1}\ge 1.
$
and $s_\pm=-a_\pm/(w'_\pm \sqrt{|b|})$.
}
\label{bw-table-E-ranges}
\end{center}
\end{table}

\paragraph{The case of $a<0$}
If $a<0$ on one side of the junction surface then \gl{sl-matching-495-500} implies
with
$
\sign(L) = \sign(\dot w)
$
the sign for $L$, the coordinate time derivative of the surface radius, which was
left undefined in \gl{sl-matching-404}.
For the case that $a_+$ and $a_-$ are negative this condition must hold on both sides
and hence a matching is only possible if
\be
\sign(\dot w_+) = \sign(\dot w_-) \quad \mbox{ for } a_+,a_-<0.
\markas{bw-eq-constraints-1000}
\ee

\section{The time-component of the Lanczos equation}
\label{bw-sec-time-lanczos}
\marginhelp{bw-sec-time-lanczos}

So far we have only considered matching of the metric and of the angular components
of the extrinsic curvature of the junction surface. The remaining matching condition comes from
the time-component of the extrinsic curvature \gl{m-ic-2000}, which contains a second
order time derivative of the junction coordinate radius, or equivalently a first-order time derivative
of the lapse function.

Rewriting the time-component of the extrinsic curvature in terms of our variable $u=N/k$ yields
\be
-K^{\mu \nu} u_\mu u_\nu
= \frac{j}{\sqrt{u^2-1}} \frac{\rmd u}{\rmd t}+j\sqrt{u^2-1}\frac{\dot l}{l} + u \frac{N'}{N},
\markas{sl-matching-700}
\ee
where the factor $N'/N$ is independent of the junction surface motion.
Taking the coordinate time derivative of \gl{sl-matching-403} and using
\gl{sl-matching-150} and \gl{sl-matching-405} we obtain
\be
\frac{j_\pm L}{\sqrt{u_\pm^2-1}} \frac{\rmd u_\pm}{\rmd t}
=-\frac{1}{2E} \frac{\rmd b}{\rmd t} -\frac{\rmd E}{\rmd t} \frac 1E wK^\theta_{\mp \theta}
-z,
\markas{sl-matching-800}
\ee
where (note that $(\rmd/\rmd t)f(t,\alpha(t)R)=u\dot f + j\sqrt{u^2-1} f'$)
\[
\flcommand
z \de u \frac{\rmd w'}{\rmd t} + j \sqrt{u^2-1}\frac{\rmd \dot w}{\rmd t}
=ju\sqrt{u^2-1}\{ w'' + \ddot w  \} +u^2(w')^\bullet +(u^2-1)(\dot w)'
.
\]
Differentiating \gl{sl-matching-400} and using \gl{m-gc-800} with $u^a{}_{|a}=2L/w$ yields
\[
\frac{\rmd E}{\rmd t}
=-\kappa L\left( \frac{\rho_s}{2} + p_s \right)
+\kappa \frac w 2 [T_{\mu \nu} u^\mu n^\nu] 
, 
\]
which expresses the coordinate time derivative of $E$.
Substituting for the first term in \gl{sl-matching-700}
allows us to evaluate the remaining junction
condition $[K^{\mu \nu}u_\mu u_\nu]=-\kappa(\rho_s/2+p_s)$.
The terms containing the surface pressure and density cancel each other
and we obtain
\be
0 = \frac{\kappa}{2} w[T^{\mu \nu}u_\mu n_\nu] + [z] -L\left[ j\sqrt{u^2-1}\frac{\dot l}{l}\right]
-L\left[ u\frac{N'}{N}\right].
\markas{sl-matching-1400}
\ee 
The first term can be expressed in terms of the Einstein-tensor with respect to
the original metric \gl{m-coord-100} as
\[
\flcommand
\kappa T_{\mu \nu} u^\mu n^\nu =
G_{\mu \nu} u^\mu n^\nu = ju\sqrt{u^2-1}\left( \frac{G_{tt}}{N^2}
+\frac{G_{rr}}{l^2} \right)
+(2u^2-1)\left( \frac{G_{tr}}{Nl} \right),
\]
where the relevant components of the Einstein-tensor are
given by
\begin{eqnarray}
G_{tt}&=&\frac{2N^2}{w}\left(
-w''+\dot w \frac{\dot l}{l} +\frac{1-a}{2w}
\right) \nonumber \\
G_{rr}&=&\frac{2 l^2}{w} \left(
\ddot w - w' \frac{N'}{N} + \frac{1-a}{2w}
\right) \nonumber \\
G_{tr}&=&\frac{2Nl}{w}\left(
\dot w \frac{N'}{N} -(w')^\bullet
\right), \nonumber
\end{eqnarray}
and $L$ is given by \gl{sl-matching-200}. Substituting into \gl{sl-matching-1400}
and using the relation
\[
(\dot w)' - (w')^\bullet = \frac{\dot l}{l} w' - \frac{N'}{N} \dot w
\]
shows that \gl{sl-matching-1400} is an identity, satisfied {\em for all} spherically
symmetric junctions between solutions of the Einstein-field equations
if the geometric matching conditions (\gl{O3-geom-cond-1} and
\gl{O3-geom-cond-2}) together with the angular component of the Lanczos
equation \gl{sl-matching-300} are satisfied. While it was well-known that for certain cases
the time-component of the Lanczos equation is identically satisfied (e.g. \cite{blau,hajicek}),
it seems to be a new result for the generic spherically symmetric case.

It was suggested that for the matching of FLRW models the time-component of
the Lanczos equation determines the pressure \cite{berezin}. In light of
the above result this cannot be the case and one needs to supplement the
model with an equation of state for the surface-matter.

\section{Matching of FLRW sections}
\label{sec-FLRW-matching}
\marginhelp{sec-FLRW-matching}

We want to turn our attention now to the special case of the matching
of two distinct FLRW regions. Such junctions are encountered in cosmological
models which approximate universes containing many FLRW domains
(multidomain universes). The most prominent example is Linde's
Chaotic Inflation scenario \cite{linde1, linde2}.

Junctions of this type have been studied in \cite{berezin,sakai2}. Our treatment will
serve as an illustration for the introduced method and as a source for numerical examples. 
Here it is not our aim to investigate physical processes which could lead the to creation of a ``bubble''
and we refer the interested reader to the vast literature (see, e.g.,
\cite{berezin,coleman-luccia,jensen-steinhardt,lyth-stewart,amendola}).
Instead we want to focus on the generic geometrical and mathematical aspects.

\subsection{FLRW models and their parametrization}

The metric of FLRW models can be written in the form
\be
\rmd s^2 = - N^2(t)\rmd t^2 +l^2(t) \{ \rmd r^2 + f^2(r) \rmd \Omega^2\} ,
\markas{FLRW-para-100}
\ee
where $l(t)$ is the scale factor, $N(t)$ the so-called lapse function, $\rmd \Omega$ the
line-element on the two-dimensional unit-sphere, and
\[
f(r) = \left\{
\begin{array}{ll}
\sin(r) &\mbox{for closed models}\\
r&\mbox{for flat models}\\
\sinh(r)&\mbox{for open models}
\end{array}
\right. .
\]
Note that the FLRW metric \gl{FLRW-para-100} has the same form as the general metric for spherical
symmetric spaces \gl{m-coord-100}, but with
$l' =0, N'/N=0$ and $\dot f=0$.

The evolution of FLRW models is described by the Friedmann equation - the dynamic part of the Einstein-Field
equations -
\be
\left( \frac{\dot l}{l} \right)^2 - \frac{\kappa \rho + \Lambda}{3} = - \frac{\zeta}{l^2},
\markas{Friedmann}
\ee
where $\zeta=0,+1,-1$ for flat, closed, and open models, respectively, and
a dot indicates the derivative with respect to {\em proper} time $t$, \iek
$\dot l \de \frac{1}{N} \frac{\rmd l}{\rmd t}$.
The matter is described by an energy-momentum tensor of perfect fluid type. The
unit tangent vectors to the fluid flow lines are given by\footnote{
In the new coordinate system which is continuous at the junction the
components take the form
$v^\mu = \frac 1N \delta^\mu_t - \frac{\dot \alpha R}{\alpha} \delta^\mu_R$.}
\[
v^\mu = \frac 1 N \delta^\mu_t,
\]
and the energy-momentum tensor describing the comoving perfect fluid takes the form
$
T^{\mu \nu} = (\rho+p)v^\mu v^\nu + p g^{\mu \nu},
$
where $\rho$ is the energy density and $p$ the pressure. The matter evolution is then
described by the energy-conservation equation
$
\dot \rho + (\rho +p) 3H =0,
$
where $H\de \dot l/l$ is the Hubble parameter.

We restrict ourself to models with a $\gamma$-law equation of state, i.e., models in which
energy density $\rho$ and pressure $p$ are related by\footnote{
The case $\gamma=0$ gives an effective cosmological constant. We can exclude this
case here because the cosmological constant is included separately.
}
\[
p=(\gamma-1)\rho
\qquad \gamma \in (2/3, 2 ].
\]
In this case $\chi_{\gamma} \de \frac{\kappa}{3} \rho l^{3 \gamma}$ is a constant of
motion. This allows us to
eliminate the energy density $\rho$ from the Friedmann equation \gl{Friedmann}, so that the
evolution of the scale factor $l$ is described in terms of the constants of motion by
\be
H^2=\chi_{\gamma} l^{-3 \gamma} + \frac 13 \Lambda - \frac{\zeta}{l^2}.
\markas{Friedmann2}
\ee

Junctions are often used as models for transition regions between two
almost-FLRW regions. Underlying is the assumption that an initially
small transition region remains small.

\subsection{Comoving junction surface}

Let us first investigate whether there could be a comoving junction surface
for two FLRW models with $\gamma$-equation of state.
This case is uniquely identified by $\dot \alpha_\pm = 0$.

From the geometric matching condition
(specialized to the FLRW metric) \gl{O3-geom-cond-1} we find
\be
l_+ = \underbrace{\frac{f_-(\alpha_-)}{f_+(\alpha_+)}}_{\rm const.} l_-,
\markas{FLRW-comoving-100}
\ee
where the first factor is now time-independent.

Let us assume we are given a solution to the Friedmann equation with $\gamma=\gamma_-, \chi=\chi_-$,
$\Lambda=\Lambda_-$, and $\zeta=\zeta_-$.
The question is now, whether there are constants $\gamma_+, \Lambda_+$, $\chi_{\gamma_+}$ and $\zeta_+$
such that $l_+ = \lambda l_-$ is a solution for constant $\lambda>0$. Both solutions would have the same
Hubble parameter $H \de \frac{\dot l_+}{l_+}=\frac{\dot l_-}{l_-}$. Substituting in  each
case from the Friedmann equation \gl{Friedmann2} gives
\[
\chi_{\gamma_-} l_-^{-3 \gamma_-} + \frac 13 \Lambda_- - \frac{\zeta_-}{l_-^2}
= \chi_{\gamma_+} l_-^{-3 \gamma_+} \lambda^{-3 \gamma_+} + \frac 13 \Lambda_+ - \frac{\zeta_+}{l_-^2}
\frac{1}{\lambda^2},
\]
which has to be satisfied for all values of $l_-$.
On both sides all terms contain different powers of $l$
(note that $-3\gamma \in (-2,6]$). In order that both sides
contain the same powers of $l$ we need
$\gamma_- = \gamma_+$.
Comparing the coefficients gives then $\lambda=1, \zeta_+=\zeta_-, \Lambda_+=\Lambda_-$, and
$\chi_{\gamma_+}=\chi_{\gamma_-}$. Hence the solutions are identical.

We conclude that if the inside and outside of the
bubble are evolving according to the Friedmann equation \gl{Friedmann} with
a $\gamma$-law equation of state then no non-trivial comoving junction is
possible.
We note that this result follows alone from the geometric matching
condition \gl{O3-geom-cond-2} --
it does not depend on the presence of a surface layer.

\subsection{Matching of FLRW regions with surface-layer}

The FLRW metric \gl{FLRW-para-100} implies
$w=l(t)f(r)$ and by taking proper-time and radial derivatives
we derive
$
\dot w = Hw
\quad \mbox{ and } \quad
w' = \frac{\rmd f}{\rmd r}.
$
The component of the energy-momentum tensor which is needed to evaluate
the surface-matter evolution according to \gl{m-gc-800} is easily found to
be (for completeness we include $k$, which is set to unity)
\[
\left[ T^{\mu \nu} u_\mu n_\nu \right]
=\left[ \frac{N^2\dot \alpha l(\rho + p)}{k^2} \right].
\]
We proceed now by expressing all quantities related to the metric and  its
derivatives ($a,b,w',\dot w$ etc.) in terms of FLRW model quantities.
With $(\rmd f/\rmd r)^2 = 1-\zeta f^2$ we obtain the expressions
$
a
\refeq{Friedmann} 1-\frac{\kappa \rho +\Lambda}{3} w^2.
$

\begin{figure}[ht]
\begin{center}
\includegraphics[height=6cm]{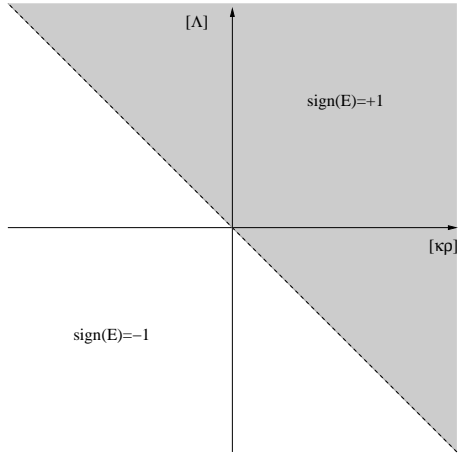}
\label{bw-fig-flrw-L-r-plane}

\caption{Points in the $\Lambda-\kappa \rho$-plane
for which a matching requires a positive/negative surface-energy density.
Only certain matchings are possible if there can only be a positive surface energy density.}
\end{center}
\end{figure}

First we want to examine which kind of bubbles could exist if there
can only be a positive surface-energy density on the junction surface, i.e., $E>0$.
For reasonably small bubbles (such that the circimference increases with the
radial coordinate)
we have
$
a_+>0, a_- > 0, and w'>0.
$
We find from table \ref{bw-table-E-ranges} that in this case
\[
\sign(E) = -\sign(b)=\sign([\kappa \rho +\Lambda]),
\]
and hence junctions are only possible if the inside FLRW region has
a smaller total\footnote{The cosmological constant represents the
vacuum energy density.} energy density than the outside region. This is illustrated in
figure \ref{bw-fig-flrw-L-r-plane}.

\subsection{The closed to (inflating) open junction}
To understand the behaviour of the junction surface
it is instructive to consider a particularly simple example for which
the evolution equations are known.
One such example is the junction between a non-inflating closed geometry with radiation inside
($\Lambda_-=0, \zeta_-=+1$)
and an inflating empty open geometry outside ($\chi_+=0, \zeta_+=-1$).

For these cases the Friedmann equation \gl{Friedmann2} is easily integrated and one finds
the well-known solutions
\be
l_-(\tau_-) = \sqrt{2 \tau_- \chi_- - \tau_-^2}
\qquad
l_+(\tau_+) = \sqrt{\frac{3}{\Lambda_+}} \sinh\left(\sqrt{\frac{\Lambda_+}{3}} \tau_+\right),
\ee
where $\tau_+$ and $\tau_-$ are the proper times along fluid flow-lines
outside and inside, respectively.
For the inside model the scale factor $l_-$ grows until it reaches a maximum at $\tau_-=\chi_-$,
and then declines until it reaches zero at $\tau_-=2\chi$. On the contrary the outside model
expands exponentially forever.

\begin{figure}[ht]
\begin{center}
\includegraphics[height=6cm]{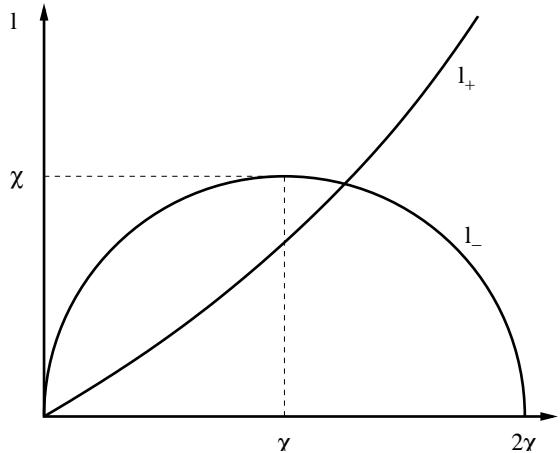}
\caption{Evolution of the scale factor in a closed non-inflating FLRW model ($l_-$) and
in an empty open inflating model ($l_+$).}
\label{bw-fig-oc-scale-factors}
\end{center}
\end{figure}

Since for each time $t$ the proper time measured along the junction surface is always
less or equal
to the proper time in the inside and outside model ($u_\pm\ge 1$) it is clear that
such a boundary can only exist for a finite proper time measured along the junction ---
the (timelike) junction surface must be `terminated' at some time.

There are four possible solutions. Firstly, it is possible that the junction surface exists
forever (in terms of the proper time) in the outer region while the proper time along the
junction surface is bounded. In our formalism this corresponds to a non-integrable
divergence in the lapse function $N_+=u_+$ for the outer region.

Secondly, the junction surface can contract to a point such that the inner
region is eliminated.
This case is characterized by $\alpha_+$, $\alpha_-$ and $w$ approaching zero at
some finite time.

\begin{figure}[ht]
\begin{center}
\includegraphics[height=3cm]{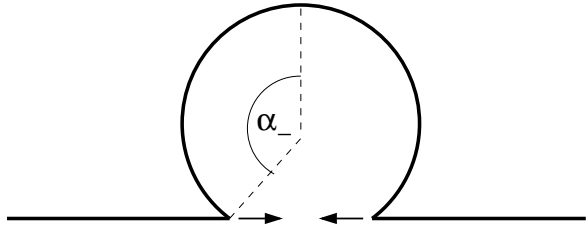}
\caption{The closed inside geometry can detach from the open outside
geometry.  This happens when $\alpha_-\rightarrow \pi$ within a finite time.}
\label{bw-fig-oc-detaching}
\end{center}
\end{figure}

Thirdly, the closed surface might detach from the open geometry --- the birth of a child universe.
In this case the radial coordinate for the closed geometry $\alpha_-$ approaches $\pi$, while
$\alpha_+$ and $w$ vanish (see figure \ref{bw-fig-oc-detaching}).

As a last option the outside region might be eliminated. In the case of a closed outside
geometry this is surely a possible solution, but in the cases of flat and open outside
geometries this generally requires that the junction surface turns spacelike\footnote{
One might speculate that for a closed inside geometry there has to be a finite volume
and hence the junction surface has to turn backwards in time and become timelike again.}.

In our formalism such a behaviour would yield a diverging, but integrable, lapse function.
At the singularity we reach the speed of light and our formalism breaks down. Nevertheless,
of the three options, to become super luminal,  to continue at the speed of light, or to
decelerate, the first one seems most convincing, also with view on the results of
section \ref{bw-subsec-expansion}.

Generally, one of these cases has to occur before we reach the singularity in the inside
region as can be seen from the following argument. Let us assume that for physical
reasons only positive surface energies are allowed. Since the outside geometry is
open we have $w'_+>0$. As the closed inside geometry approaches the `big crunch'
singularity the energy density grows without bound. Hence $b=w^2(\kappa \rho_- - \Lambda_+)$
has to become positive at some stage during the contraction phase. However, from
table \ref{bw-table-E-ranges} one can see that there is no solution possible with $w'_+>0$ and
$b>0$ if $a_+$ and $a_-$ are positive. Let us note that when the inside region is contracting
we have $\dot w_-<0$ and $\dot w_+>0$. Hence according to \gl{bw-eq-constraints-1000}
$a_+$ and $a_-$ cannot be both negative. If $a_+$ is positive (and $a_-$ negative) then
\gl{bw-eq-constraints-600} implies that the surface energy density is negative, which is
in contradiction with our assumption.
If on the other hand $a_+<0$ then this implies $L>0$ and hence the proper surface radius
would increase. This just helps driving $a_-=1-\kappa \rho_- w^2/3$
closer to zero, which eventually has to turn
negative due to the diverging energy density. Again we reach a point where no solution is possible
without negative surface energies.

Note that if one allows negative surface energies then the above argument shows that
if the junction starts with a positive surface-energy density then at
some point the junction must have a vanishing surface-energy density.

Figure \ref{bw-numeric-graph16} shows the results of a numerical integration of this particular
model. It appears as if the speed of light is reached within a finite time (integrable
lapse functions) on both sides.
This strongly suggests to us that the junction turned spacelike.
Note that this happens even far before the
inner closed region enters the contracting phase.

With this example we want to emphasize that there are junctions which are possible
initially, but which evolve to some singular point. Numerical studies have shown
that this is rather common.

\subsection{Vacuum bubbles}
As a simple case vacuum bubbles ($\rho+p=0$ on both sides)
have frequently appeared in the literature
\cite{berezin,sakai2}. For these cases the surface energy-momentum
tensor takes the form $S^\mu{}_\nu=-\rho_s \delta^\mu{}_\nu$ and
it follows from \gl{m-gc-800} that $\rho_s$ must be constant \cite{berezin}.

Assuming \cite{sakai2} $\kappa \rho + \Lambda=$const. on each side one
can integrate the evolution equation for the angular metric component
\gl{sl-matching-404} to obtain
\be
w(t) = w_0 \cosh(t/w_0),
\ee
where $t$ is the proper time along the junction and
\be
\flcommand
w_0=\frac{12 \kappa \rho_s}{\sqrt{(4(\kappa(\rho_+ + \rho-) +\Lambda_+ + \Lambda_-)
+3 \kappa^2 \rho_s^2)^2 -64(\kappa \rho_+ + \Lambda_+)(\kappa \rho_- + \Lambda_-)}},
\ee
which agrees with the findings in \cite{sakai2,berezin}.

However, this does not yet establish the actual motion of the junction surface
since the relation of the proper time (along the junction)
to the coordinate time is unknown --- we need the lapse function which is
given by \gl{sl-matching-410}. The proper time along fluid flow lines
(which is proportional to coordinate time) is then given
by $\int N \rmd t$.

\section{Numerical Results}
\label{bw-sec-numerics}

A computer program has been written to integrate the evolution equations
for several FLRW junctions
numerically. To achieve better accuracy around the singularities a variable step-width
was used. All examples given here are for positive surface-energy densities. Cases with
negative surface energy can easily be constructed by exchanging the inside and outside
region.
The graphs on the following pages illustrate the results and will be discussed one-by-one below.

\begin{sidewaysfigure}
\centering  
\subfigure[Coordinate radius of the junction surface on each side]{
\includegraphics[width=2.5in]{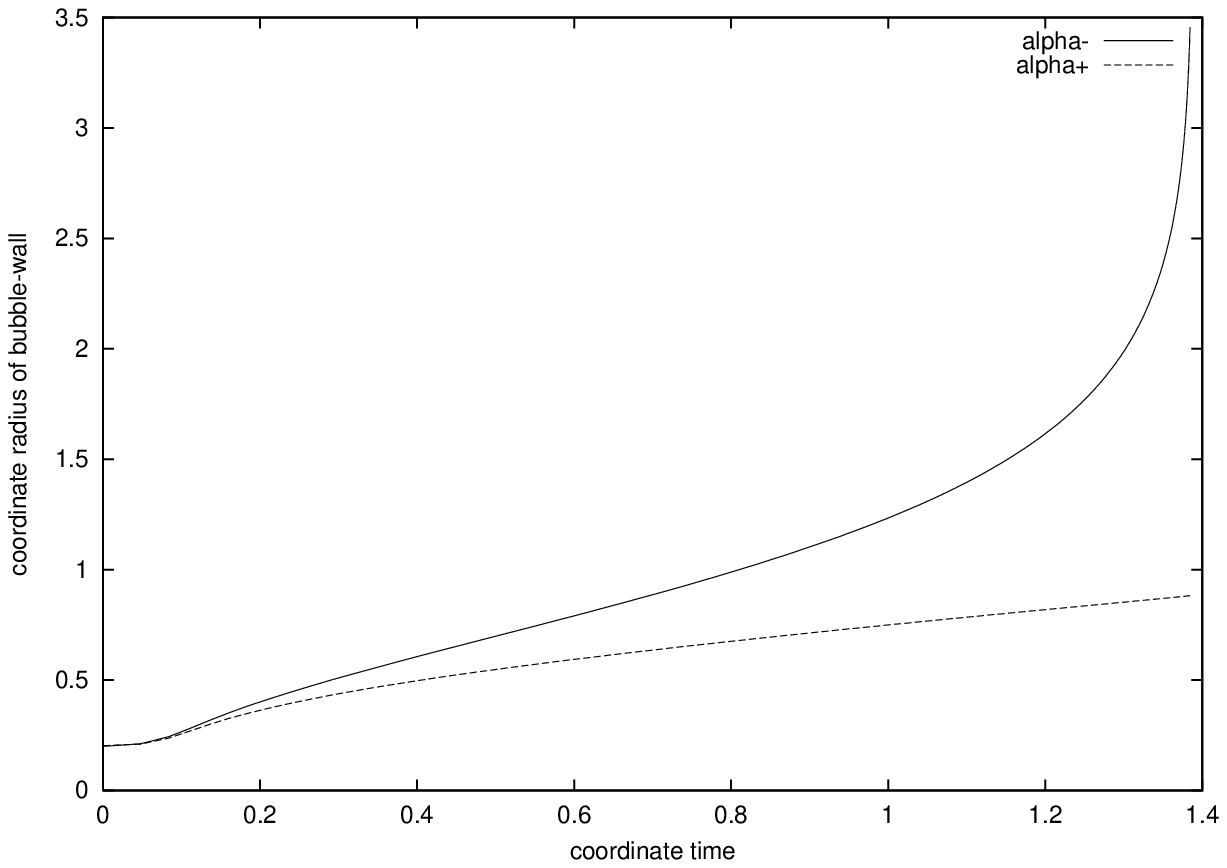}}
\hspace{.2cm}
\subfigure[Proper speed of the junction surface on each side.]{
\includegraphics[width=2.5in]{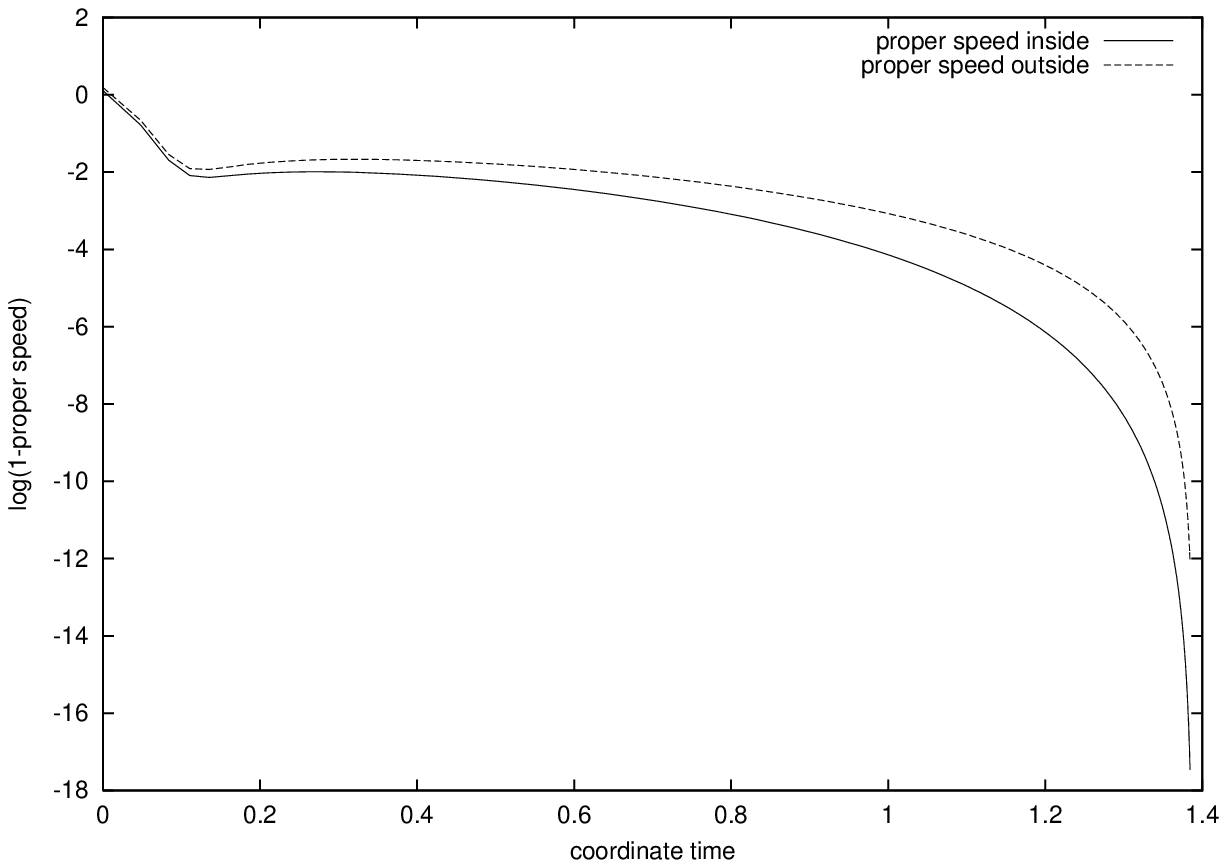}}
\hspace{.2cm}
\subfigure[Lapse function $N=u$ on each side]{
\includegraphics[width=2.5in]{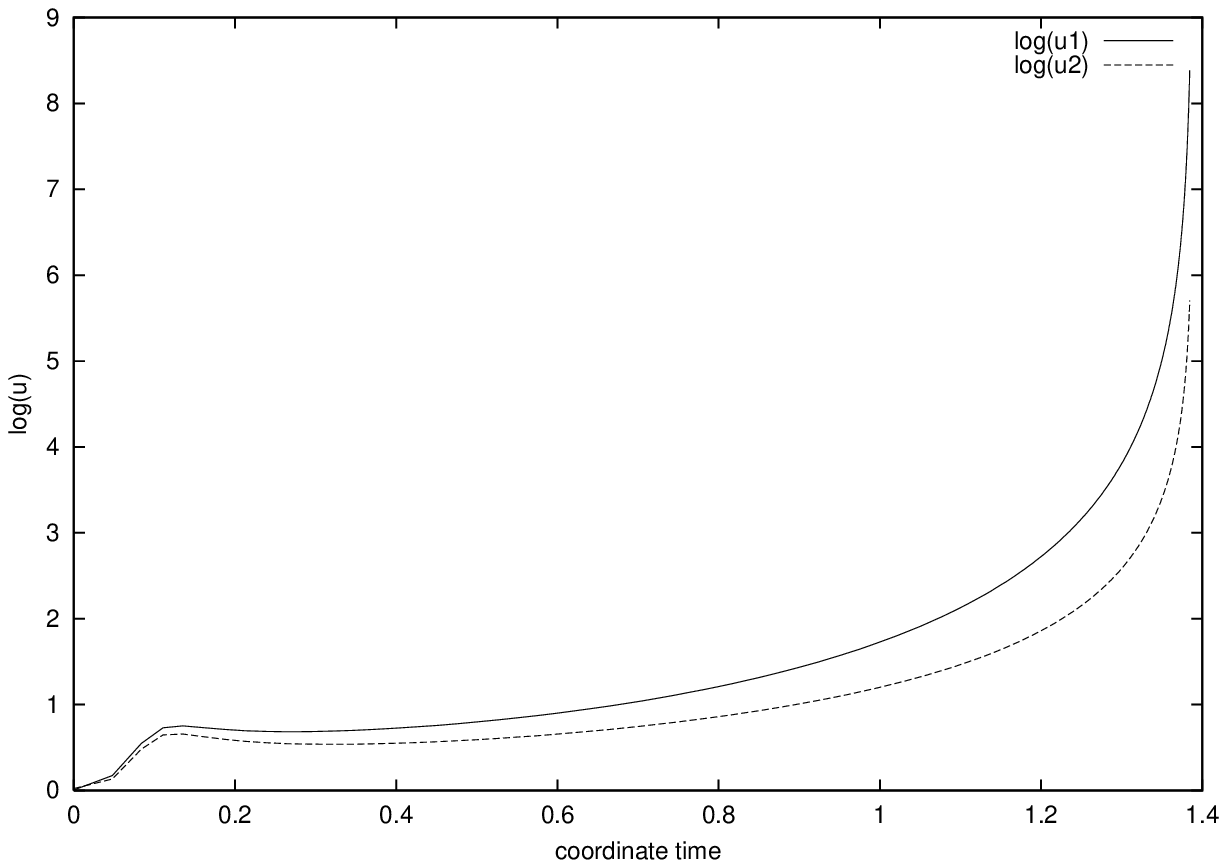}}
\\
\subfigure[Proper times on both sides]{
 \includegraphics[width=2.5in]{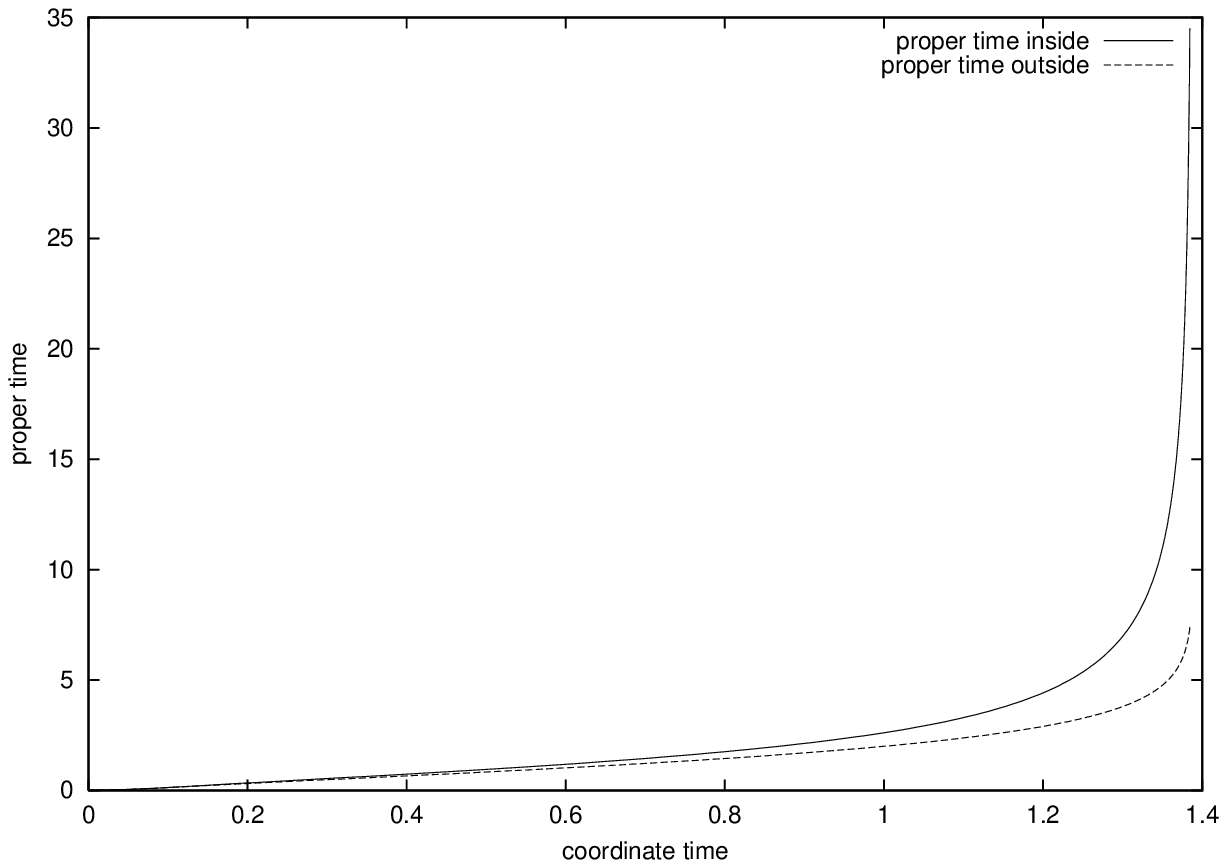}}
\hspace{.2cm} 
\subfigure[Surface radius ($w$).]{
 \includegraphics[width=2.5in]{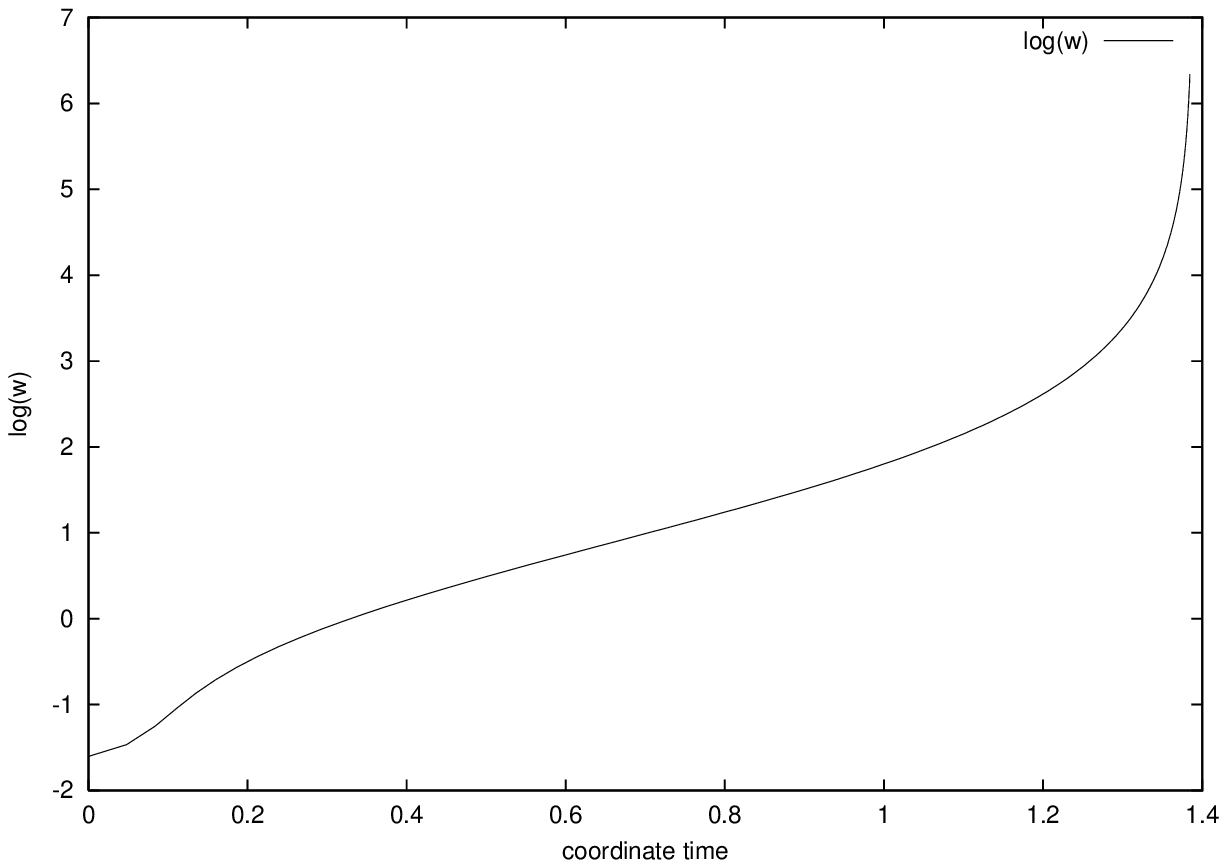}}
\hspace{.2cm} 
\subfigure[$E$, which is related to the surface-matter energy density $\rho_s$
by $E=\kappa w \rho_s/2$.]{
 \includegraphics[width=2.5in]{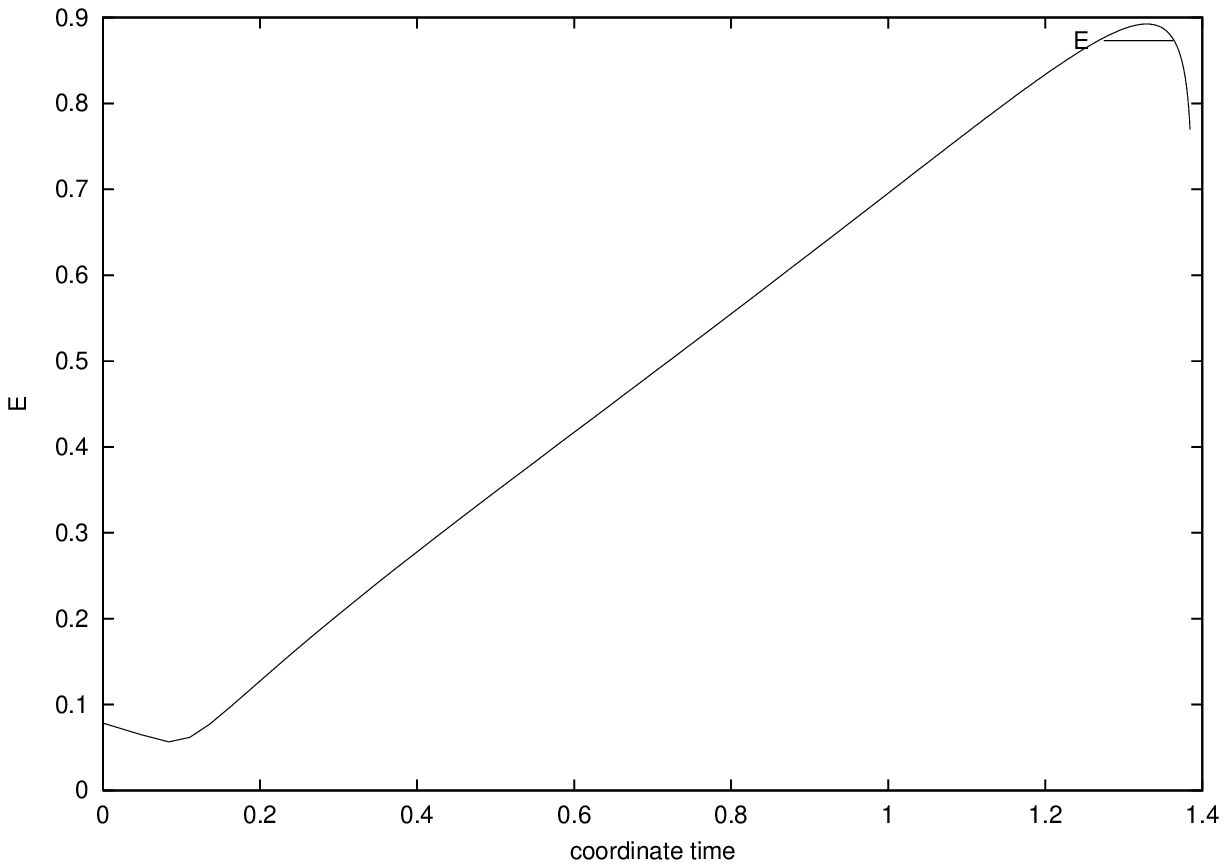}}
\caption{Evolution of a junction between an open (inside) and an inflating closed (outside)
geometry.
(Parameters:$\gamma_s=1,
\zeta_+=+1,		 \zeta_-=-1,
\chi_+=5,	 \chi_-=2,
\gamma_+=0.7,	 \gamma_-=1.9,
\Lambda_+=2,	 \Lambda_-=0$;
 Initial values:
$\rho_s=0.031,
\alpha_+=0.20272, \alpha_-=0.2,
l_+=1,		 l_-=1,
\rho_+=0.597,	 \rho_-=0.239$.
)
Note that here $u$ and $w$ are plotted logarithmically.
}
\label{bw-numeric-graph13}
\end{sidewaysfigure}

\paragraph{\it Open inside, inflating closed geometry outside}
Figure \ref{bw-numeric-graph13}
shows such an example. After some time the surface radius starts to diverge (note
the logarithmic scaling) while $E=\kappa w \rho_s/2$ does not approach zero
(hence close to the divergence $\rho_s \propto 1/w$). The proper times on both sides seem to be diverging,
which is in agreement with the results from subsection \ref{bw-subsec-expansion}.
Note that the inner and outer regions have a rather unusual equation of state with
$\gamma_+=0.7$ and $\gamma_-=1.9$ --- in this case it is really this choice of the
equations of state which makes the energy-momentum tensor contribution
$[T_{\mu \nu} u^\mu n^\nu]$ positive for small values of $E$ (see subsection \ref{bw-subsec-expansion}).

\begin{sidewaysfigure}
\centering  
\subfigure[Coordinate radius of the junction surface on each side]{
\includegraphics[width=2.5in]{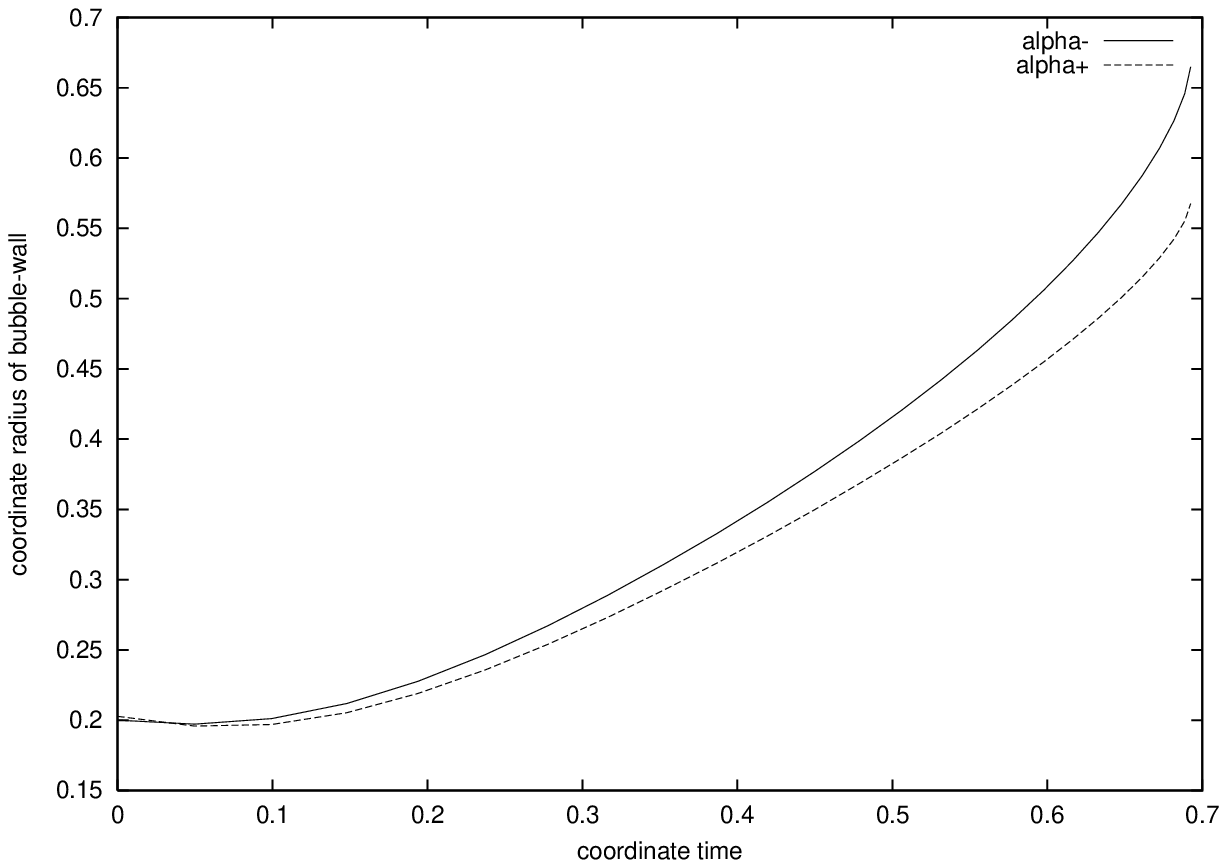}}
\hspace{.2cm}
\subfigure[Proper speed of the junction surface on each side.]{
\includegraphics[width=2.5in]{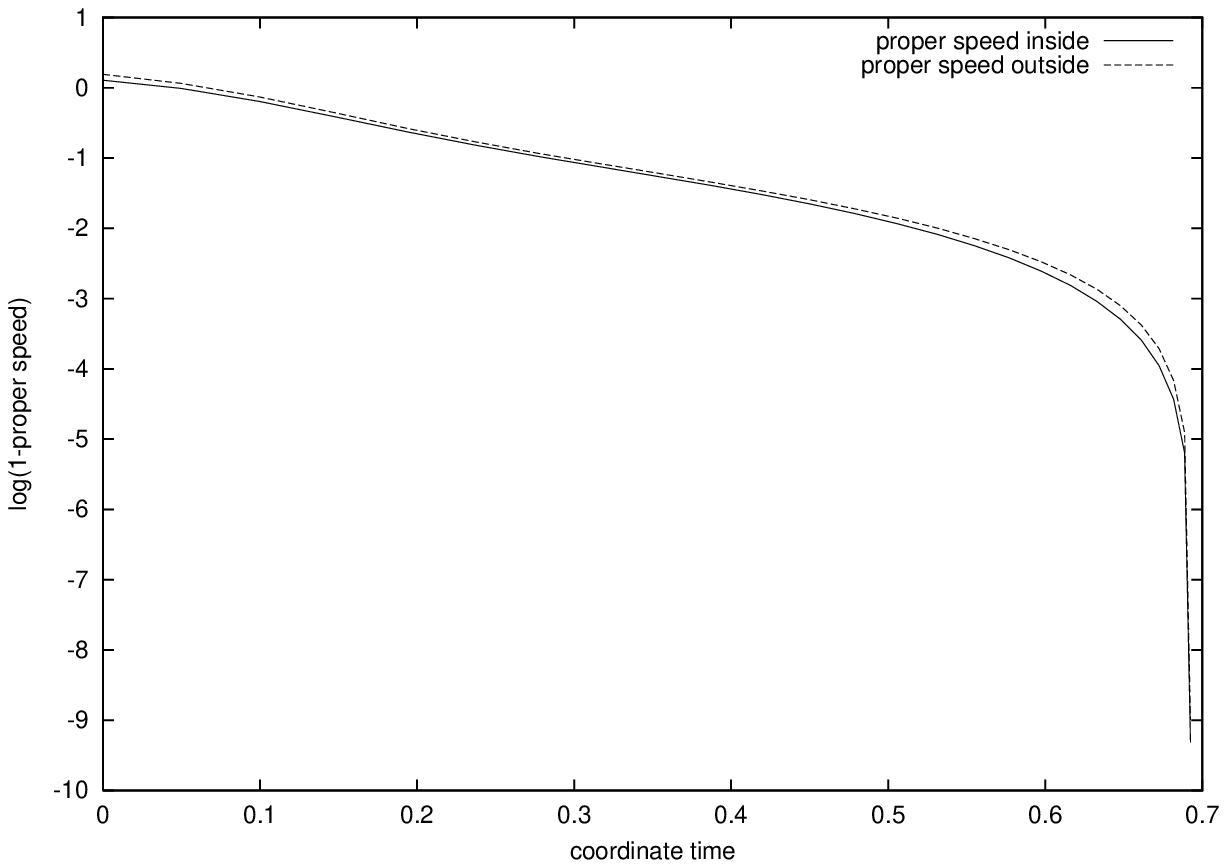}}
\hspace{.2cm}
\subfigure[Lapse function $N=u$ on each side]{
\includegraphics[width=2.5in]{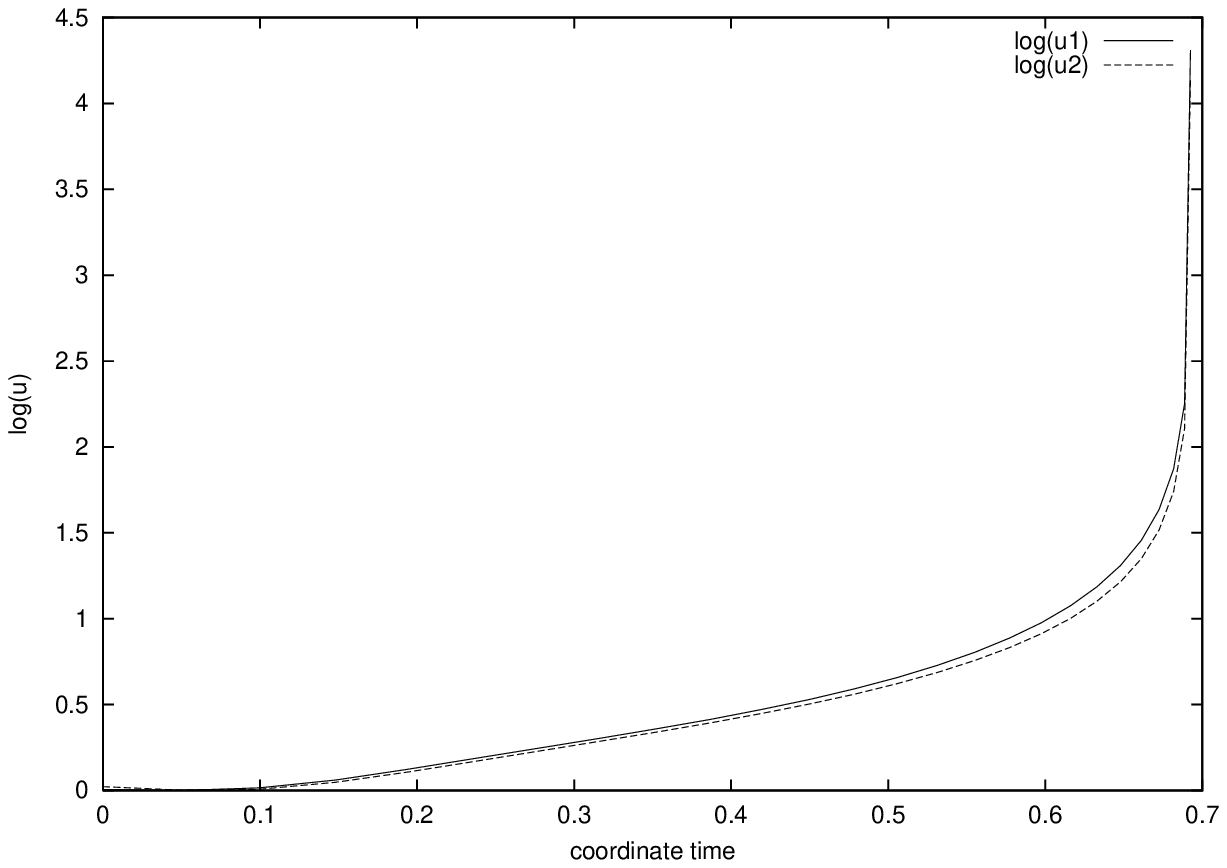}}
\\
\subfigure[Proper times on both sides]{
 \includegraphics[width=2.5in]{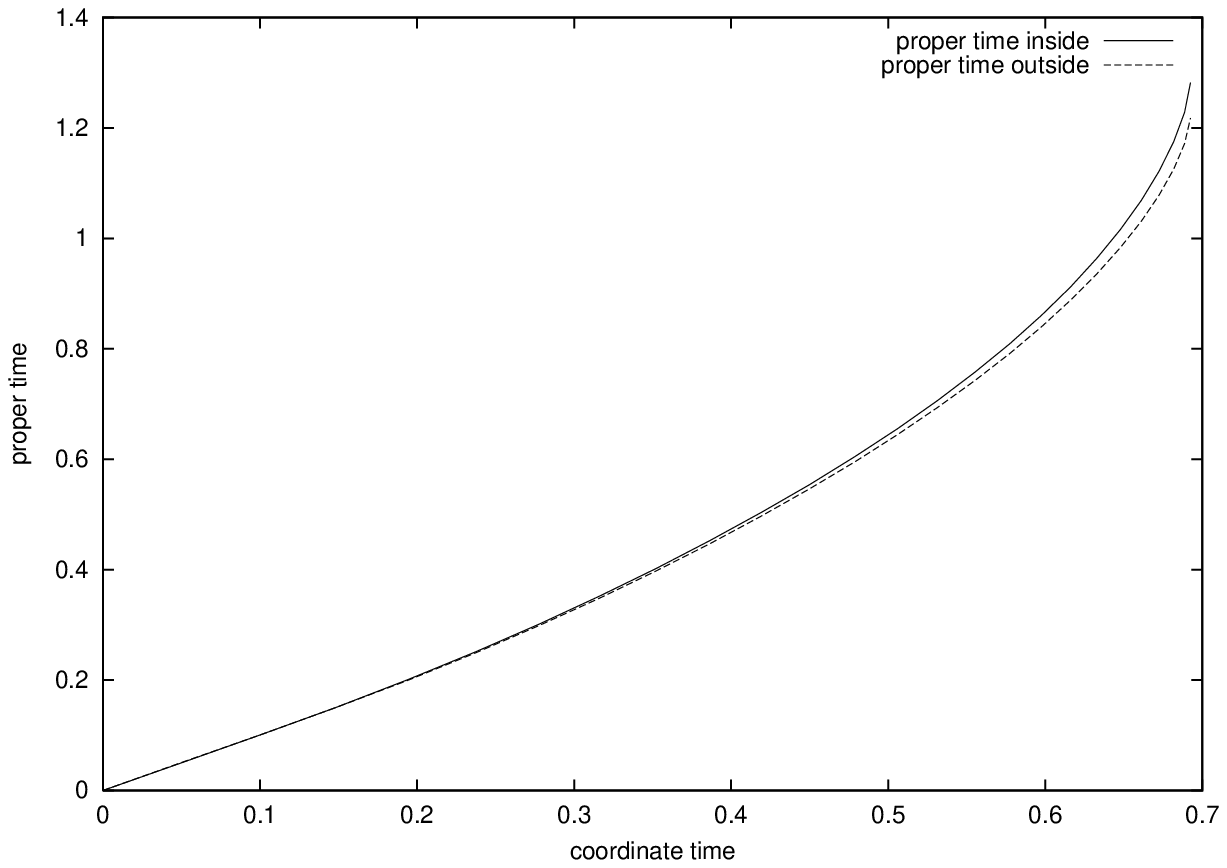}}
\hspace{.2cm} 
\subfigure[Surface radius ($w$).]{
 \includegraphics[width=2.5in]{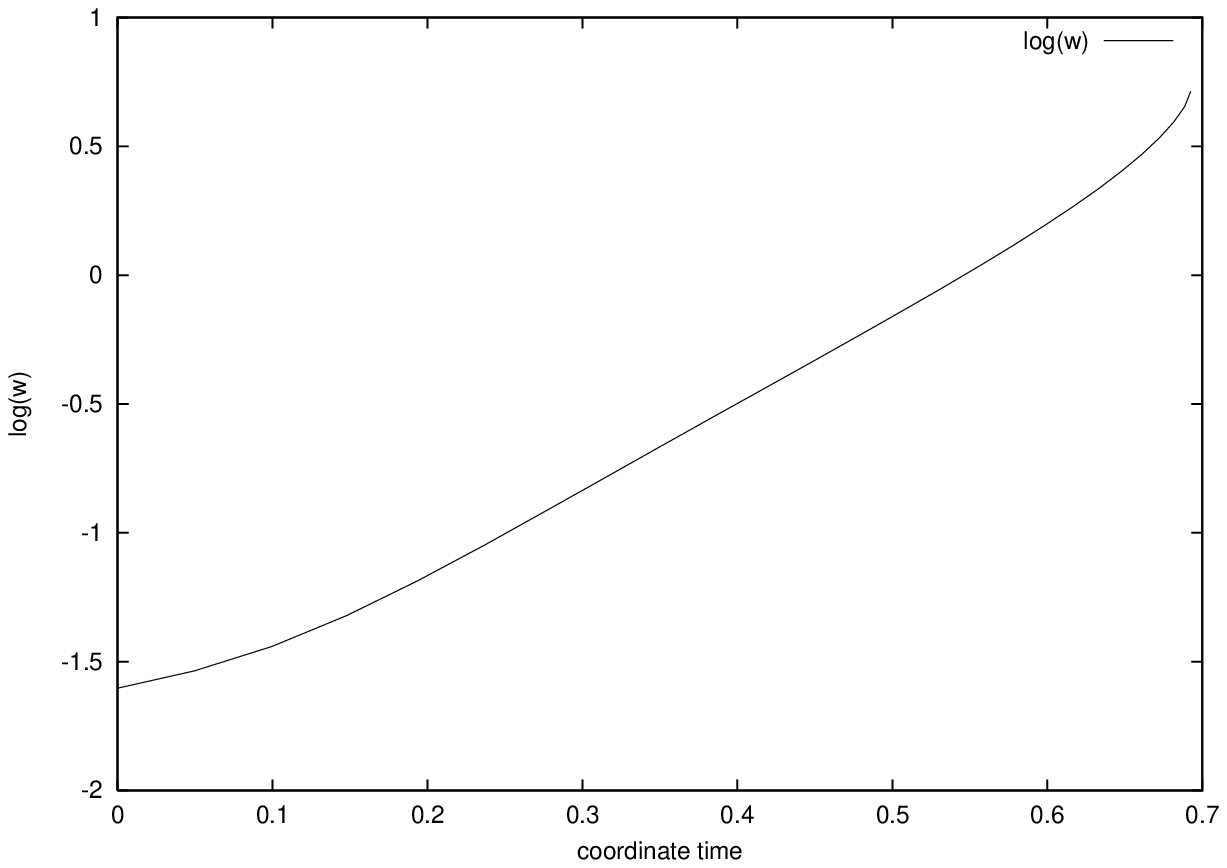}}
\hspace{.2cm} 
\subfigure[$E$, which is related to the surface-matter energy density $\rho_s$
by $E=\kappa w \rho_s/2$.]{
 \includegraphics[width=2.5in]{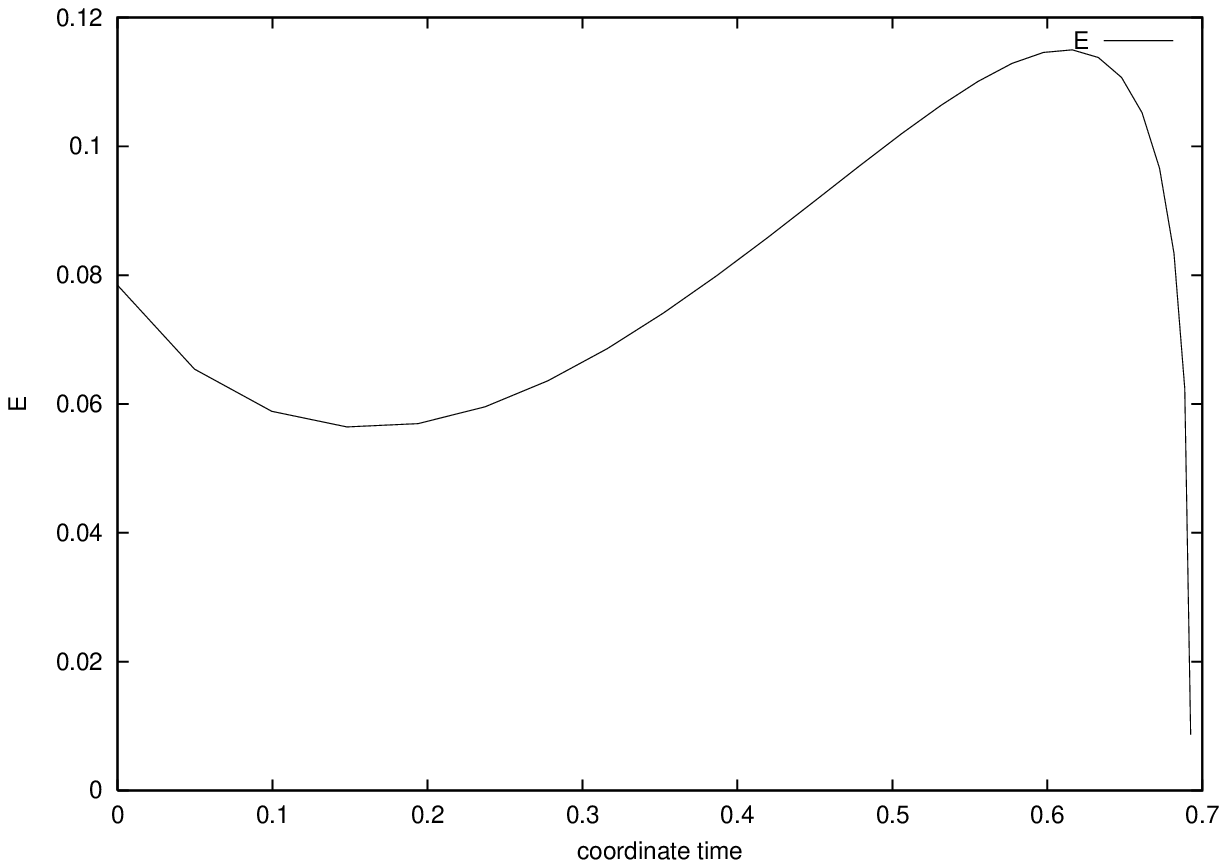}}
\caption{Evolution of a junction between an open (inside) and an inflating closed (outside)
geometry. The lapse functions are integrable and the speed
of light is reached within a finite time.
(Parameters:$\gamma_s=1,
\zeta_+=+1,		 \zeta_-=-1,
\chi_+=5,	 \chi_-=2,
\gamma_+=1,	 \gamma_-=1,
\Lambda_+=2,	 \Lambda_-=0$;
 Initial values:
$\rho_s=0.031,
\alpha_+=0.20272, \alpha_-=0.2,
l_+=1,		 l_-=1,
\rho_+=0.597,	 \rho_-=0.239$.
)
Note that here $u$ and $w$ are plotted logarithmically.
}
\label{bw-numeric-graph14}
\end{sidewaysfigure}

Figure \ref{bw-numeric-graph14} shows the evolution of the same initial situation, but with
different equations of state (dust on both sides). This seems to change the
sign of the energy-momentum contribution $[T_{\mu \nu} u^\mu n^\nu]$ for small values of $E$,
which now approaches zero within a finite time. In fact, it can be verified that
close to the singular point $t_0$ we have as expected $E \propto \sqrt[3]{t_0-t}$.
As predicted in subsection \ref{bw-subsec-expansion} the lapse functions appear to
be integrable and the proper times do not diverge. The junction surface seems
to reach the speed of light within a finite time.

Clearly, our formalism breaks down at this point. However, one
could argue that after reaching the speed of light within a finite time, one should expect the
junction to turn spacelike.

\begin{sidewaysfigure}
\centering
\subfigure[Coordinate radius of the junction surface on each side]{
\includegraphics[width=2.5in]{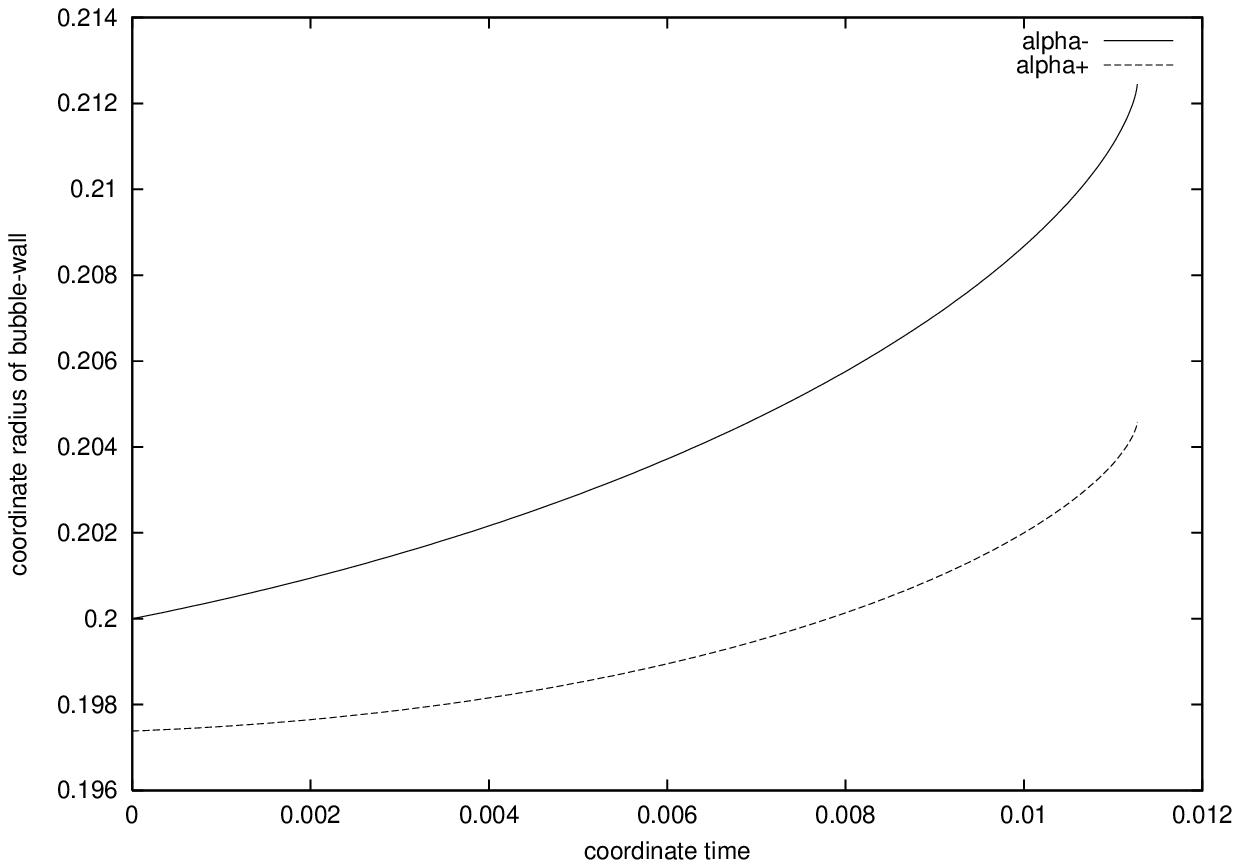}}
\hspace{.2cm}
\subfigure[Proper speed of the junction surface on each side.]{
\includegraphics[width=2.5in]{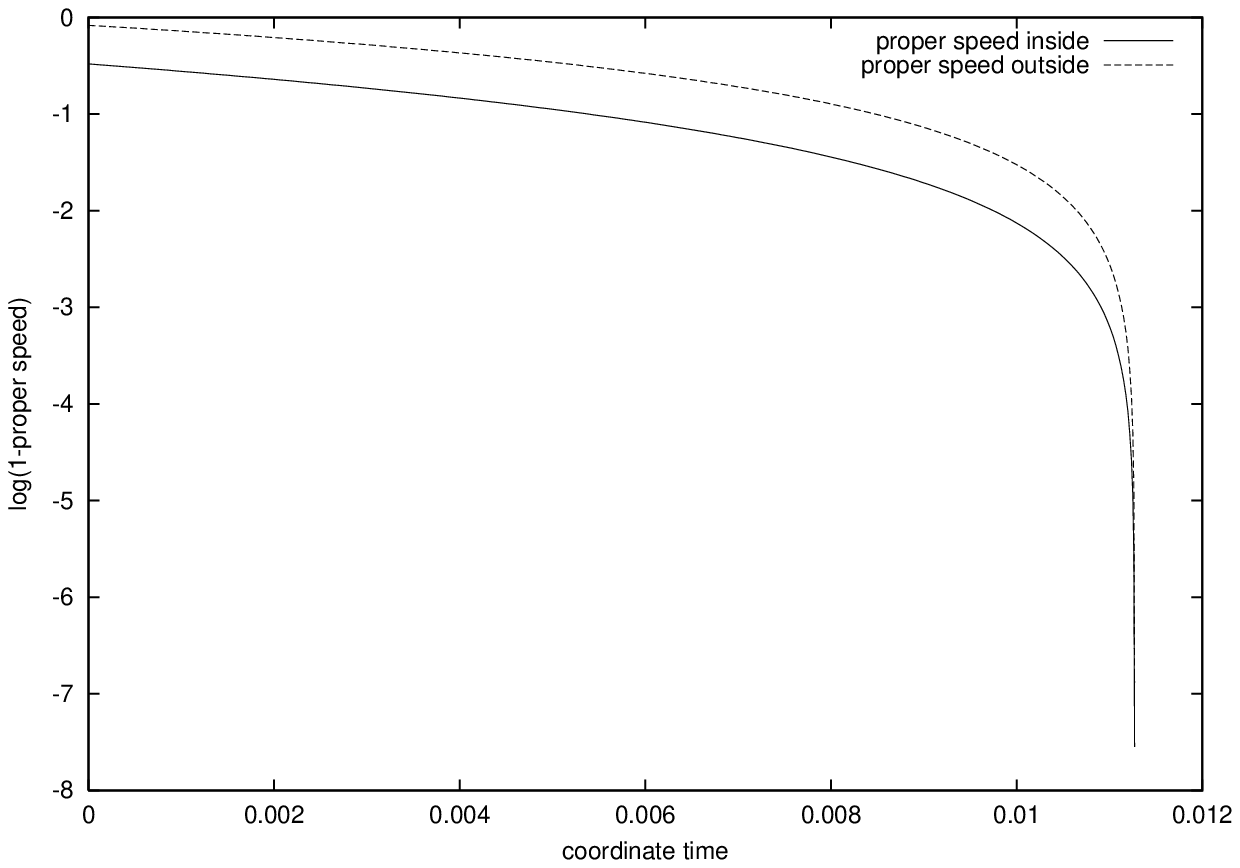}}
\hspace{.2cm}
\subfigure[Lapse function $N=u$ on each side]{
\includegraphics[width=2.5in]{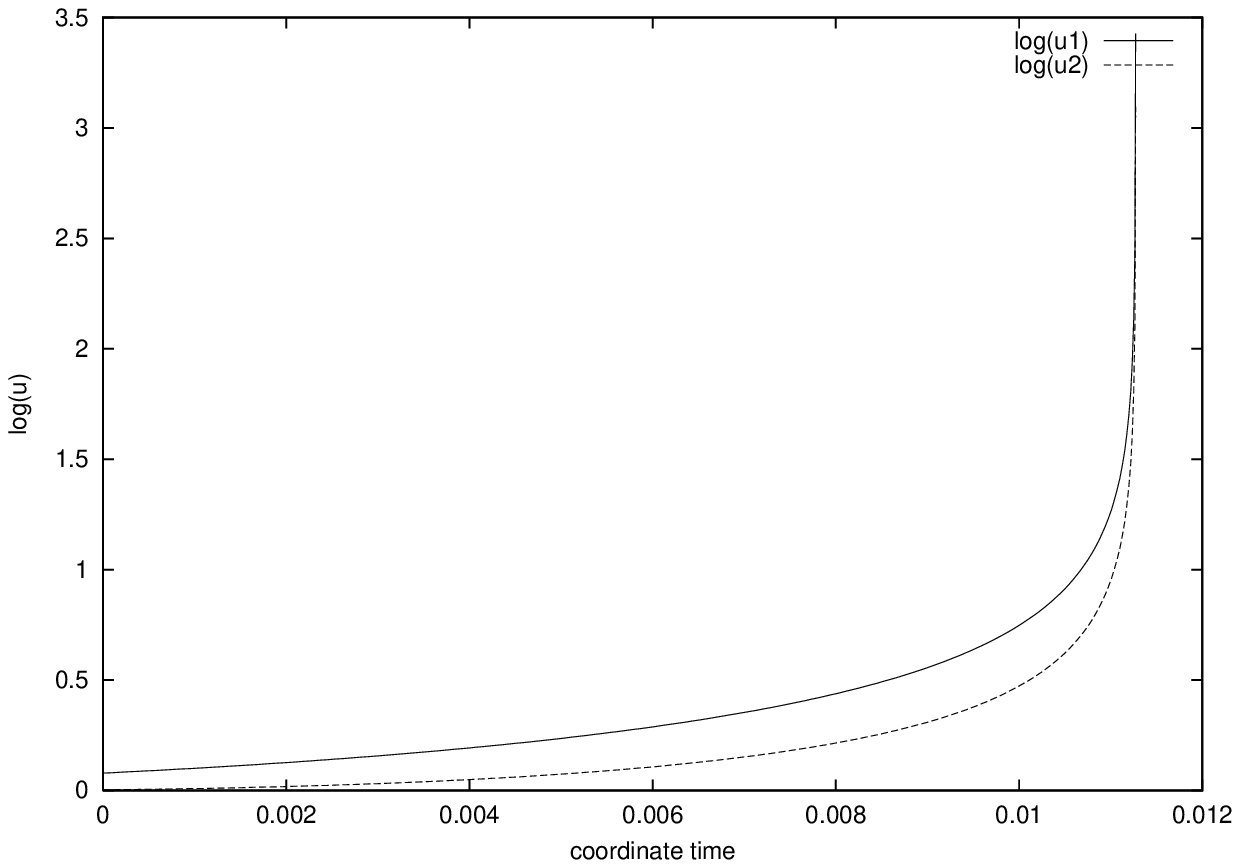}}
\\
\subfigure[Proper times on both sides]{
 \includegraphics[width=2.5in]{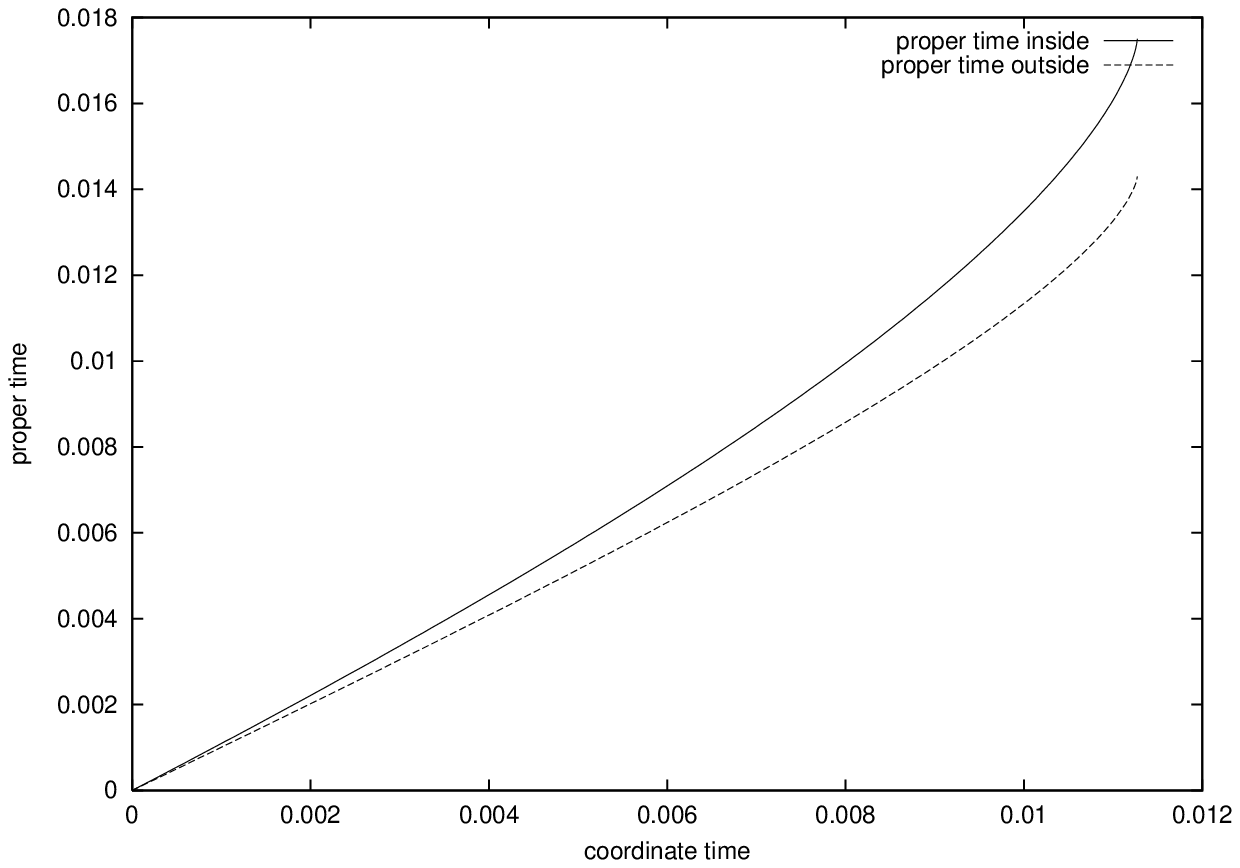}}
\hspace{.2cm}
\subfigure[Surface radius ($w$).]{
 \includegraphics[width=2.5in]{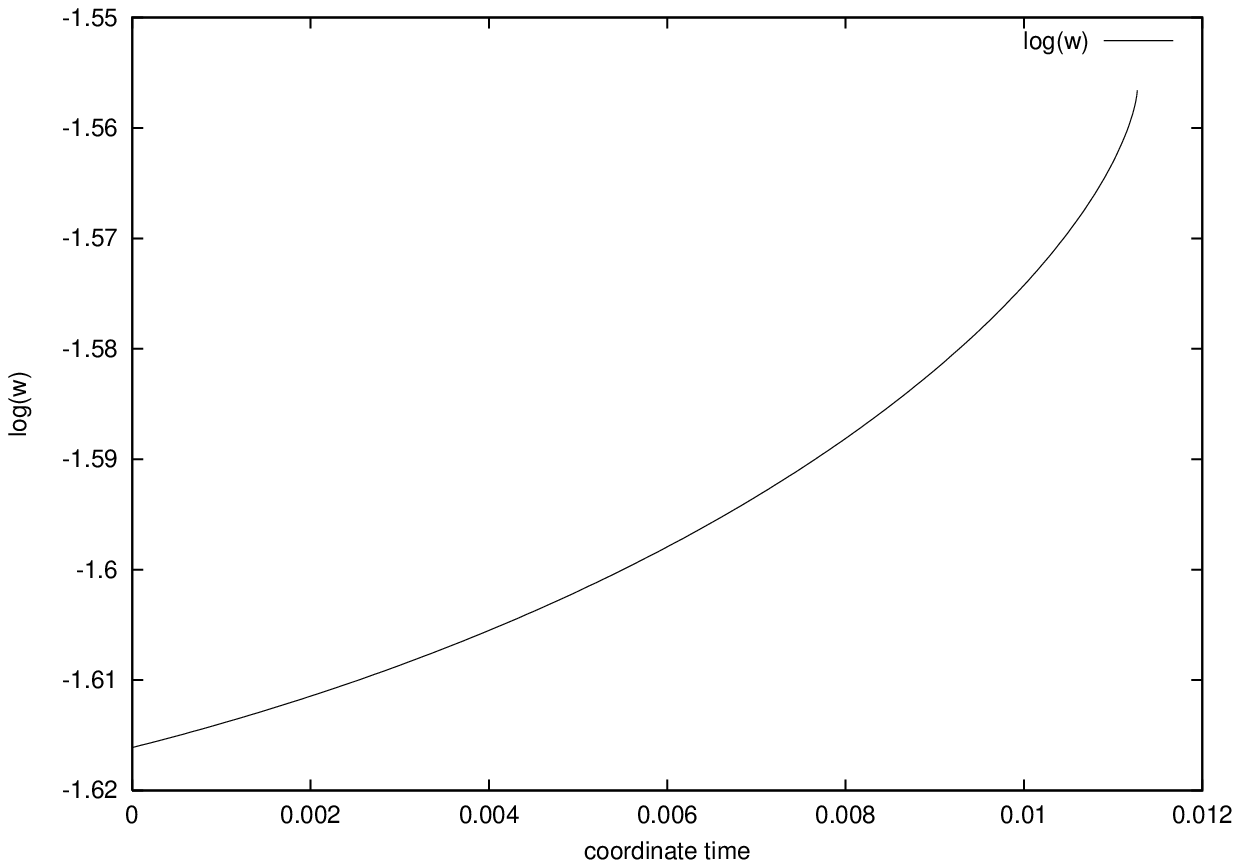}}
\hspace{.2cm}
\subfigure[$E$, which is related to the surface-matter energy density $\rho_s$
by $E=\kappa w \rho_s/2$.]{
 \includegraphics[width=2.5in]{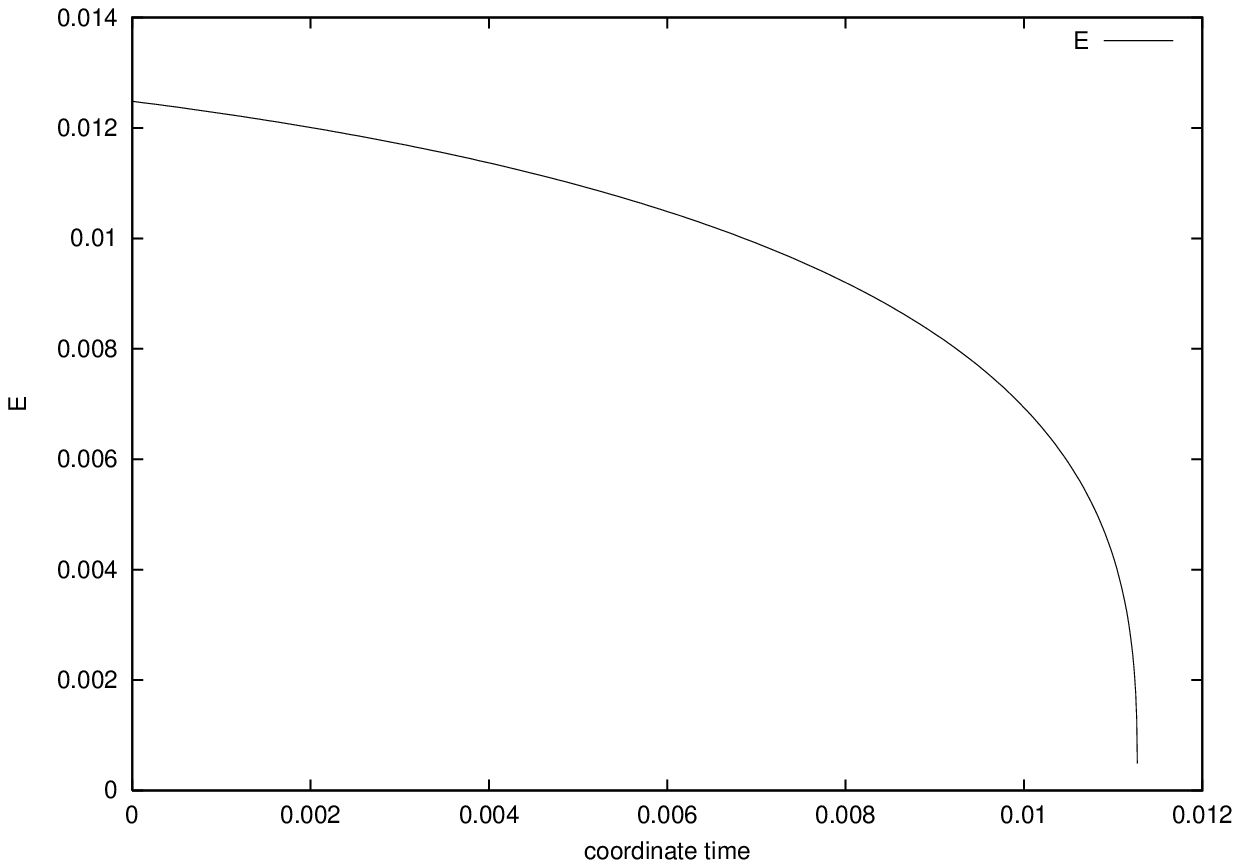}}
\caption{Evolution of a junction between a closed (inside) and an inflating open (outside)
geometry. The inside has a radiation equation of state, while the outside is an inflating
dust model.
(Parameters:
$\gamma_s=1,
\zeta_+=-1,		 \zeta_-=+1,
\chi_+=0,	 \chi_-=1,
\gamma_+=1,	 \gamma_-=4/3,
\Lambda_+=5,	 \Lambda_-=0$;
 Initial values:
$\rho_s=0.005,
\alpha_+=0.19739, \alpha_-=0.2,
l_+=1,		 l_-=1,
\rho_+=0,	 \rho_-=0.119$.
)
Note that here $u$ and $w$ are plotted logarithmically.
}
\label{bw-numeric-graph16}
\end{sidewaysfigure}

\paragraph{\it Closed inside, inflating open outside}
Figure \ref{bw-numeric-graph16} shows such a situation with a radiation equation of
state for the inside region ($\gamma_-=4/3$). This is the example discussed above.

\section{Conclusion}

We developed a formalism for the treatment of timelike junctions between spherically symmetric
solutions of the Einstein-field equation, which is based on the Lanczos equation and
the Israel junction conditions. We introduce new coordinates such that two conditions
are satisfied:
Firstly, all coordinates are continuous at the junction surface, and secondly, the
junction surface becomes a surface of constant `radial' coordinate. In this approach
the actual movement of the junction surface is absorbed into the metric, of which the
transverse components are discontinuous at the junction surface.

We evaluate the junction conditions and re-discover with \gl{sl-matching-402} and
\gl{sl-matching-403} well-known relations between the extrinsic curvatures, the
surface layer energy density, and the rate of change of the surface radius of the
junction surface.
It should be pointed out that these results follow without using the time-component  of the
Lanczos equation. As it was shown in section \ref{bw-sec-time-lanczos},
for all spherically symmetric cases this remaining
equation is in fact an identity. This was known for special cases, but it appears
to be a new result in this general form.

The behaviour for small values of $E=\kappa w \rho_s/2$ has been investigated.
It was shown that for certain cases $E$ is driven to zero within a finite
coordinate and proper time. At such a point our formalism breaks down. Nevertheless,
we want to speculate here that in such cases the junction really turns spacelike.
This can be seen as an inadequacy of the thin wall formulation in such situations ---
a causal propagation of a discontinuity should not exceed the speed of light.
We suggest that in such cases the spatial extent of the transition region is not negligible.

The developed formalism gives us two sources for constraints on possible junctions.
Firstly the time derivative of the surface radius is given by the quadratic
equation \gl{sl-matching-404}. Demanding that real solutions to this equation must
exist directly restricts the possible values of the
surface energy density for a particular junction (see figure \ref{bw-fig4new}).
Secondly, in our approach physical solutions must have a lapse function
which is greater than or equal to unity. The resulting restrictions depend
on the metric components on each side of the junction --- they either
determine the sign of the derivative of the proper surface radius, or they
restrict the possible surface-energy densities (see figure \ref{bw-fig-E-ranges}).
For the latter case the allowed ranges for $E= \kappa w \rho_s/2$ have been
given explicitly.

For the special case of junctions between FLRW models with $\gamma$-equation
of state it was shown that alone on
geometrical grounds there can be no comoving junction surface --- whether with or
without surface layer.

As a particularly simple and well-known case, `vacuum bubbles' were discussed.
and the results agree with the literature.
A particularly interesting model, the junction between an empty, open, inflating FLRW
region outside and a radiation dominated closed FLRW model inside, has been investigated
in more detail. The inside region re-collapses after some finite proper time and hence
the junction surface has to be terminated. Besides a disappearance or a detachment of the
closed inner region we suggest that the junction can turn
spacelike --- an effective disappearance
of the outer region.

This and other examples have been integrated numerically. It was observed that
many models seem to reach the speed of light within a finite {\em proper} time,
in accordance with the predictions from section \ref{bw-subsec-expansion}.
Since a spacelike junction violates causality, we suggest that a breakdown in the
thin-wall approximation must have occurred.

Our results show that the thin-wall treatment of timelike junctions (without
the presence of scalar fields) is on a {\em mathematical}
sound level. Nevertheless, in many cases the junction surface reaches a singular point
within a finite proper time. We believe that in these cases the thin-wall is not a {\em physically}
acceptable approximation.



\begin{thebibliography}{30}
\addcontentsline{toc}{section}{References}
\bibitem{spergel} Spergel D N {\it et al} 2003 {\it Astrophys. J.} ({\it Preprint} astro-ph/0302209)
\bibitem{guth} Guth A H 1981 \PR D {\bf 23} 347
\bibitem{linde1} Linde A D 1983 \PL B {\bf 129} 177
\bibitem{linde2} Linde A D 1990 \textit{Particle Physics and Inflationary
Cosmology} (Chur:Harwood Academic Publishers)
\bibitem{lanczos} Lanczos C 1922 {\it Phys. Zeits.} {\bf 23} 539; and
1924 {\it Ann. der Phys.} {\bf 74} 518
\bibitem{dautcourt} Dautcourt R 1964 {\it Math. Nachr.} {\bf 27} 277
\bibitem{israel} Israel W 1966 \NC B {\bf 44} 4349
\bibitem{blau} Blau S K, Guendelman E I and Guth A H 1987 \PR D {\bf 35} 1747
\bibitem{berezin} Berezin V A, Kuzmin V A and Tkachev I I 1987 \PR D {\bf 36} 2919
\bibitem{lake} Lake K 1987 in {\it Vth Brazilian School of Cosmology and
Gravitation}, Ed. Novello, World Scientific, Singapore, pp1-82
\bibitem{hajicek} H\'aj\'\i\v{c}ek P and Bi\v{c}\'ak J 1997 \PR D {\bf 56}
4706-19 ({\it Preprint} gr-qc/9706022)
\bibitem{sakai2} Sakai N and Maeda K 1994
\PR D {\bf 50} 5425-5428 ({\it Preprint} gr-qc/9311024)
\bibitem{misner-thorne-wheeler} Misner C W, Thorne K S and Wheeler J A 1973
{\it Gravitation}, W. H. Freeman and Company, San Francisco
\bibitem{coleman-luccia} Coleman S and De Luccia F 1980 \PR D {\bf 21} 3305
\bibitem{jensen-steinhardt} Jensen L G and Steinhardt P J 1984 {\it Nucl. Phys.} B {\bf 237} 176-188
\bibitem{lyth-stewart} Lyth D H and Stewart E D 1994 {\it Preprint} hep-ph/9408324
\bibitem{amendola} Amendola L {\it et al} 1996 {\it Preprint} astro-ph/9610038

\end{thebibliography}
\end{document}